\documentclass[prb,groupedaddress,nofootinbib,showpacs,twocolumn,floatfix]{revtex4}
\usepackage{revsymb}
\usepackage{bm}
\usepackage{graphicx}
\usepackage{amsfonts}
\usepackage{natbib}
\usepackage{bibentry}
\usepackage{hyperref}
\usepackage{gtdeleq}

\usepackage{color}
\definecolor{IITred}{rgb}{0.5,0.05,0.05}
\definecolor{TrueBlue}{rgb}{0.05,0.05,0.5}
\definecolor{DrabGreen}{rgb}{0.05,0.5,0.05}

\newcommand{\cf}{{\em cf.\ }}
\newcommand{\eg}{{\em e.g.}}
\newcommand{\ie}{{\em i.e.}}

\newcommand{\gev}{\hbox{ GeV}}

\newcommand{\mev}{\hbox{ MeV}}

\newcommand{\tev}{\hbox{ TeV}}

\newcommand{\fm}{\hbox{ fm}}

\newcommand{\Eqn}[1]{Eq.~(\ref{#1})}

\newcommand{\cfrac}[2]{{\textstyle \frac{#1}{#2}}}

\newcommand{\gtimes}{\times}

\newcommand{\cgg}{\ensuremath{\mathrm{SU(3)}_{\mathrm{c}}}}

\newcommand{\ewgg}{\ensuremath{\mathrm{SU(2)_L}\gtimes\mathrm{U(1)}_Y}}

\newcommand{\D}{\ensuremath{\mathcal{D}}}
\newcommand{\tr}[1]{\mathop{\mathrm{tr}}#1}
\newcommand{\blambda}{\bm{\lambda}}
\newcommand{\alphas}{\ensuremath{\alpha_{\mathrm{s}}}}

\newcommand{\citentry}[1]{%
\par\addvspace{0.5em plus 0.05em minus 0.05em}%
\sloppy
\noindent\textbf{\onlinecite{#1}.}~\bibentry{#1}%
\par\addvspace{0.5em plus 0.05em minus 0.05em}
}

\begin{document}

\nobibliography{QCDRL}

\preprint{FERMILAB-PUB-10/040-T}

\title[QCD Resource Letter]{Resource Letter QCD-1: Quantum Chromodynamics}



\author{Andreas S. Kronfeld}
\email[E-mail address: ]{ask@fnal.gov}
%
\author{Chris Quigg}
\email[E-mail address: ]{quigg@fnal.gov}
\affiliation{Theoretical Physics Department,
Fermi National Accelerator Laboratory,
P.O. Box 500, Batavia, Illinois 60510 USA}


\date{\today}

\begin{abstract}
	This Resource Letter provides a guide to the literature on Quantum
	Chromodynamics (QCD), the relativistic quantum field theory of the
	strong interactions.
	Journal articles, books, and other documents are cited for the
	following topics:
    quarks and color,
    the parton model,
    Yang-Mills theory,
    experimental evidence for color,
    QCD as a color gauge theory,
    asymptotic freedom, 
    QCD for heavy hadrons,
    QCD on the lattice,
    the QCD vacuum,
    pictures of quark confinement,
    early and modern applications of perturbative QCD,
    the determination of the strong coupling and quark masses,
    QCD and the hadron spectrum,
    hadron decays,
    the quark-gluon plasma,
    the strong nuclear interaction,
    and
	QCD's role in nuclear physics.
	
	The letter {E} after an item indicates elementary level or material
	of general interest to persons becoming informed in the field.  The
	letter {I}, for intermediate level, indicates material of a somewhat
	more specialized nature, and the letter {A} indicates rather
	specialized or advanced material.
\end{abstract}
%
%
%
%
%
%
%
%
%
%
%
%
\pacs{12.38.-t, 24.85.+p \hfill FERMILAB-PUB-10/040-T}

\maketitle

\tableofcontents

\section{Introduction}

Quantum chromodynamics (QCD) is a remarkably simple, successful, and rich
theory of the strong interactions.
The theory provides a dynamical basis for the \textit{quark-model}
description of the \textit{hadrons}, the strongly interacting particles
such as protons and pions that are accessible for direct laboratory
study.
Interactions among the quarks are mediated by vector force particles
called \textit{gluons}, which themselves experience strong interactions.
The nuclear force that binds protons and neutrons together in atomic
nuclei emerges from the interactions among quarks and gluons.

QCD describes a wealth of physical phenomena, from the structure of
nuclei to the inner workings of neutron stars and the cross sections
for the highest-energy elementary-particle collisions.
The QCD literature is correspondingly extraordinarily vast.
To arrive at a manageable number of cited papers in this 
Resource Letter, we have chosen works that should be useful to 
professors and students planning a course for classroom or independent 
study. 
We have included some classic contributions and all elementary
presentations of which we are aware.
For more advanced material we favor works that provide ambitious 
students an entr\'{e}e to the contemporary literature. 
These include well-established highly-cited review articles (because a
search of literature citing a review is a gateway to newer topics)
and more modern treatments that portray the state of the art and 
document the preceding literature well.
Wherever possible, we give links to digital versions of the articles we
cite.
Many published articles are available in electronic form through the
World Wide Web sites of individual journals, or through the e-print 
archive; see Appendix~\ref{app:links}.

From the time of its development in the 1950s, quantum electrodynamics
(QED), the relativistic quantum field theory of photons and electrons,
was viewed as exemplary.
In the late 1960s and early 1970s, with the development of the
electroweak theory, it became increasingly attractive to look to
relativistic quantum field theories---specifically gauge theories---for
the description of all the fundamental interactions.
\citentry{Wilczek:1998ma}
\noindent QCD represents the culmination of that search for the strong
interactions.
In some respects, it has supplanted QED as our ``most perfect'' theory.
\citentry{Wilczek:1999id}

An excellent summary of the foundations and implications of QCD is
\citentry{Nobel2004}
\noindent Some encyclopedia articles on QCD are
\citentry{Kronfeld:96}
\citentry{Sterman:2005vn}
\citentry{cqqcd}
\noindent For a book-length exposition of the wonders of QCD, see
\citentry{Wilczek08}

The rest of this Resource Letter is organized as follows.
We begin in Sec.~\ref{sec:qcd} by reviewing the basics of the theory of 
QCD, giving its Lagrangian, some essential aspects of its dynamics, and
providing a connection to earlier ideas.
In Sec.~\ref{sec:tools} we cover literature on theoretical tools for 
deriving physical consquences of the QCD Lagrangian.
Section~\ref{sec:expt} covers the most salient aspects of the 
confrontation of QCD with experimental observations and measurements.
Section~\ref{sec:beyond} situates QCD within the broader framework 
of the standard model of particle physics.
We conclude in Sec.~\ref{sec:outlook} with a brief essay on frontier 
problems in QCD.
Appendix~\ref{app:links} gives links to basic online resources.

\section{QCD}
\label{sec:qcd}

As a theory of the strong interactions, QCD describes the properties of 
hadrons.
In QCD, the familiar mesons (the pion, kaon, etc.)\ are bound states of 
\emph{quarks} and \emph{antiquarks}; the familiar baryons
(the proton, neutron, $\Delta(1232)$ resonance, etc.)\ are bound states of 
three quarks.
Just as the photon binds electric charges into atoms, the binding agent 
is the quantum of a gauge field, called the gluon.
Hadrons made of exclusively of gluons, with no need for valence quarks, 
may also exist and are called glueballs.
Properties of hadrons
are tabulated in
\citentry{Amsler:2008zzb}

In this section we begin with the Lagrangian formulation of~QCD.
Readers who are not yet familiar with the Dirac equation may wish to 
skip this mathematical discussion and head straight 
to Sec.~\ref{sec:qcd:first} for a r\'{e}sum\'e of the main themes of QCD,
to Sec.~\ref{sec:qcd:texts} for a list of textbooks, or
to Sec.~\ref{sec:qcd:parents} for resources on the ideas out of which 
the quantum field theory QCD emerged in the early 1970s. 

\subsection{A gauge theory for the strong interactions}
\label{sec:qcd:basics}

Quantum chromodynamics is the theory of strong interactions among
quarks derived from the color gauge symmetry \cgg. 
It is advantageous for many purposes to express the theory in 
Lagrangian form.
As in a classical theory, one can easily derive the equations of motion.
In a quantum theory, the Lagrangian also provides a convenient 
framework for quantization and the development of perturbation theory, 
via Feynman rules, in a Lorentz-covariant fashion.
The Lagrangian formalism lends itself particularly to the consideration
of symmetry principles and their consequences.
Invariance of the Lagrangian under a \emph{global}, \ie,
position-independent, symmetry operation implies a conservation law
through Noether's theorem.
Requiring the Lagrangian to be invariant under \emph{local}, \ie,
position-dependent, transformations demands an interacting theory, in
which spin-one force particles couple minimally to the conserved current
of the global symmetry.
Thus a global U(1) phase symmetry is related to conservation of the
electromagnetic current, and local U(1) phase symmetry underlies QED.
A symmetry used to derive a theory of interactions is called a
\emph{gauge symmetry.}

In nature, we find six \emph{flavors} of quarks, ``up,'' ``down,'' 
``charm,'' ``strange,''  ``top,'' and ``bottom.''
The electric charges of the quarks are $2e/3$ for the up, charm, and top 
flavors, and $-e/3$ for the down, strange, and bottom flavors, where 
$-e$ is the electon's charge.
The essence of the QCD Lagrangian is captured for a single flavor:
\begin{equation}
	\mathcal{L} = \bar{\psi}(i \gamma^{\mu}\D_{\mu} -m)\psi -
	\cfrac{1}{2}\tr{(G_{\mu\nu}G^{\mu\nu})}.
	\label{eq:qcdlag}
\end{equation}
The composite spinor for color-triplet quarks of mass $m$ is
\begin{equation}
	\psi = \left( 
	\begin{array}{l}
	q_{\mathrm{red}} \\ q_{\mathrm{green}} \\ q_{\mathrm{blue}}
	\end{array}
	\right),
\end{equation}
where each element $q_i$ is a four-component Dirac spinor, acted upon 
by the Dirac matrices $\gamma^\mu$.
The gauge-covariant derivative is
\begin{equation}
\D_{\mu} = \partial_{\mu} + i g B_{\mu}\;,
\end{equation}
where $g$ is the strong coupling constant and the object $B_{\mu}$ is a
three-by-three matrix in color space formed from the eight (gluon)
color gauge fields $B^{l}_{\mu}$ and the generators 
$\cfrac{1}{2}\lambda^{l}$ of \cgg\ as
\begin{equation}
B_{\mu} = \cfrac{1}{2} \bm{B}_{\mu} \cdot \blambda  =
\cfrac{1}{2}B^{l}_{\mu}\lambda^{l}\;.
\end{equation}
The gluon field-strength tensor is
\begin{eqnarray}
G_{\mu\nu} & = & \cfrac{1}{2}\bm{G}_{\mu\nu} \cdot \blambda =
\cfrac{1}{2}G^{l}_{\mu\nu}\lambda^{l} \\
 & = & (ig)^{-1}\left[\D_{\nu},\D_{\mu}\right] = 
 \partial_{\nu}B_{\mu} - \partial_{\mu}B_{\nu} +
ig\left[B_{\nu},B_{\mu}\right].
 \nonumber
\end{eqnarray}
The $\lambda$-matrices satisfy
\begin{eqnarray}
\tr{(\lambda^{l})} & = & 0, \\
\tr{(\lambda^{k}\lambda^{l})} & = & 2\delta^{kl} ,\\
\left[\lambda^{j},\lambda^{k}\right] & = & 2if^{jkl}\lambda^{l}\;,
\end{eqnarray}
and the \emph{structure constants} $f^{jkl}$ can be expressed as
\begin{equation}
    f^{jkl} = (4i)^{-1}\,
    \tr{\left\{\lambda^{l}\left[\lambda^{j},\lambda^{k}\right]\right\}} \;.
\end{equation}
The nonvanishing structure constants distinguish QCD from QED: 
QCD is a \emph{non-Abelian} gauge theory.
The gluon field-strength can be expressed in component form as
\begin{equation}
    G^{l}_{\mu\nu} = \partial_{\nu}B^{l}_{\mu} - \partial_{\mu}B^{l}_{\nu}
        + gf^{jkl}B^{j}_{\mu}B^{k}_{\nu},
\end{equation}
and the last term marks a fundamental dynamical difference between QCD 
and QED.
Via $\tr{(G_{\mu\nu}G^{\mu\nu})}$ in Eq.~(\ref{eq:qcdlag}), it leads to 
three-gluon and four-gluon interactions that have no counterpart in QED.
The gluon carries color charge and, thus, experiences strong 
interactions, whereas the neutral photon does not couple directly to other
photons.

The color matrices for the fundamental (quark) representation satisfy
\begin{equation}
    \sum_i \lambda^i_{ab}\lambda^i_{bc} = 4C_F\delta_{ac}, 
    \quad C_F = \frac{N^2 - 1}{2N},
    \label{eq:fcolorfact}
\end{equation}
while the color matrices for the adjoint (gluon) representation, 
$T^k_{ij} = -if^{ijk}$, obey
\begin{equation}
    \tr{(T^k T^l)} = \sum_{ij}f^{ijk}f^{ij\ell} = C_A \delta^{k\ell},
    \quad C_A = N.
    \label{eq:gcolorf}
\end{equation}
For QCD based on $\mathrm{SU(}N=3\mathrm{)}$ gauge symmetry, the quark 
and gluon color factors are 
\begin{equation}
    C_F = \cfrac{4}{3}, \quad C_A = 3.
    \label{eq:qgcolorf}
\end{equation}
It is sometimes advantageous to carry out calculations for general 
values of $C_F$ and $C_A$, to test the non-Abelian structure of QCD 
(see Sec.~\ref{subsec:shapes}).

Physical arguments in favor of the \cgg\ gauge theory
are collected in
\citentry{Fritzsch:1973pi}

The Lagrangian $\mathcal{L}$ in Eq.~(\ref{eq:qcdlag}) is invariant under 
the transformations
\begin{eqnarray}
    \psi(x)  & \mapsto & e^{i\omega(x)}\psi(x), \label{eq:gt:psi} \\
    \bar{\psi}(x)  & \mapsto & \bar{\psi}(x)e^{-i\omega(x)}, \\
    B_\mu(x) & \mapsto & e^{i\omega(x)}[B_\mu + (ig)^{-1}\partial_\mu]
        e^{-i\omega(x)} \label{eq:gt:B} ,
\end{eqnarray}
where the matrix 
$\omega(x)=\cfrac{1}{2}\omega^l(x)\lambda^l$ depends on the spacetime 
coordinate~$x$.
Generically the matrices $\omega(x)$ and $B_\mu(x)$ do not commute, a 
feature that again distinguishes QCD from QED.

If we recast the matter term in the Lagrangian (\ref{eq:qcdlag}) in
terms of left-handed and right-handed fermion fields,
\begin{equation}
	\mathcal{L}_{q} =
		\bar{\psi}_{\mathrm{L}}i\gamma^{\mu}\D_{\mu}\psi_{\mathrm{L}}
		+ \bar{\psi}_{\mathrm{R}}i\gamma^{\mu}\D_{\mu}\psi_{\mathrm{R}}
		-m\left(\bar{\psi}_{\mathrm{L}}\psi_{\mathrm{R}} +
		\bar{\psi}_{\mathrm{R}}\psi_{\mathrm{L}} \right),
    \label{eq:chiral}
\end{equation}
we see that it becomes highly symmetrical in the limit of vanishing
quark mass, $m \to 0$. Absent the mass term, there is no coupling
between the left-handed and right-handed quark fields,
$\psi_{\mathrm{L,R}} \equiv \cfrac{1}{2}(1 \mp \gamma_5)\psi$, and so
the Lagrangian is invariant under separate global phase transformations
on the left-handed and right-handed fields. Generalizing to the case of
$n_{\mathrm{f}}$ flavors of massless quarks, we find that the QCD
Lagrangian displays an
$\mathrm{SU(\mathit{n}_{\mathrm{f}})_L}\gtimes%
\mathrm{SU(\mathit{n}_{\mathrm{f}})_R}\gtimes%
\mathrm{U(1)_L}\gtimes\mathrm{U(1)_R}$
\emph{chiral symmetry}.
In nature, the up- and down-quark masses are very small (compared to 
the proton mass), and the strange-quark mass is also small.
Therefore, $n_{\mathrm{f}}=2$ (isospin) and $n_{\mathrm{f}}=3$ (flavor 
$\mathrm{SU}(3)$) chiral symmetries are approximate.
We return to chiral symmetries in Sec.~\ref{lightquarks}.

The U(1) factors may be rewritten $\mathrm{U}(1)_V\gtimes\mathrm{U}(1)_A$.
The vector ($V$) symmetry applies the same phase factor to left- and 
right-handed fields; it leads via Noether's theorem to a conserved 
charge, namely baryon number.
The axial-vector ($A$) symmetry applies opposite phase factors to left- 
and right-handed fields;
it is broken by certain quantum-mechanical effects, called anomalies, 
discussed in Sec.~\ref{sec:anomaly}.

\subsection{First consequences}
\label{sec:qcd:first}

As mentioned above, the three- and four-gluon interactions make the 
physics of QCD essentially different from the mathematically similar QED.
In quantum electrodynamics, an electron's charge is partially screened
by vacuum polarization of the surrounding cloud of virtual
electron-positron pairs.
The effect can be measured with a probe of wavelength $1/Q$, 
and described by a scale dependence, or running, of the fine structure 
constant $\alpha\equiv e^2/4\pi$.
Omitting charged particles other than the electron, a first-order 
calculation of the running yields
\begin{equation}
    \frac{1}{\alpha(Q)} = \frac{1}{\alpha(m_e)} - 
        \frac{2}{3\pi} \log\left(\frac{Q}{m_e}\right)\;,
\end{equation}
where $m_e$ is the electron's mass, and the formula holds for $Q>m_e$.
Note the sign of the logarithm: at larger values of $Q$, which is to 
say shorter distances, the effective charge increases.

In quantum chromodynamics, gluons can fluctuate into further 
quark-antiquark pairs, and this vacuum polarization exerts a similar 
screening effect, tending to increase the effective color charge at 
short distances.
But this tendency is overcome by \textit{antiscreening}
effects that arise from the contributions of gluon loops to the vacuum
polarization. 
The gluon loops are present because of the three-gluon and four-gluon 
vertices that arise from the non-Abelian nature of the \cgg\ symmetry.
To one-loop approximation, the strong-interaction analogue of the fine 
structure constant, $\alphas\equiv g^2/4\pi$, evolves as
\begin{equation}
	\frac{1}{\alphas(Q)} = \frac{1}{\alphas(\mu)} + 
		\frac{33-2n_{\mathrm{f}}}{6\pi}
		\log\left(\frac{Q}{\mu}\right),
	\label{eq:asevol}
\end{equation}
where $\mu$ defines the reference, or renormalization, scale. 

If the number of quark flavors $n_{\mathrm{f}} \le 16$, as it is in our
six-flavor world, then the coefficient of the $\log(Q/\mu)$ term
is \textit{positive,} and $\alphas$ decreases at large 
values of $Q$ or short distances.
This is the celebrated property of \textit{asympotic freedom,} 
announced in
\citentry{Gross:1973id}
\citentry{Politzer:1973fx}

Asymptotic freedom points to the existence of a
domain in which the strong interactions become sufficiently weak that
scattering processes can be treated reliably in perturbation theory
using techniques based on the evaluation of Feynman diagrams.  
The path to asymptotic freedom is described in the Nobel Lectures,
\citentry{Gross:2005kv}
\citentry{Politzer:2005kc}
\citentry{Wilczek:2005az}
\noindent For another view of the historical setting, see 
\citentry{'tHooft:1998xb}
\noindent Asymptotically free theories are of special interest because 
they predict behavior very close to Bjorken scaling in deeply inelastic
scattering (see Secs.~\ref{sec:qcd:partons} and~\ref{sec:DIS}).
No renormalizable field theory without non-Abelian gauge fields can be
asymptotically free:
\citentry{Coleman:1973sx}
\noindent Asymptotic freedom is thoroughly established in laboratory 
scattering experiments, as discussed in Sec.~\ref{sec:runalpharun}.

The complementary behavior of QCD in the long-distance limit, known as
\textit{infrared slavery}, points to the confinement of quarks into
color-singlet hadrons, as explained in 
\citentry{Nambu:1976ay} 
\noindent This picture leads to the crucial insight that most of the 
mass of hadrons such as the proton arises not from the masses of their 
constituents, the quarks, but from the quarks' kinetic energy and the 
energy stored in the gluon field, 
\citentry{Wilczek:1999be}
\citentry{Wilczek:2000it}
\citentry{Wilczek:2006eg}
\citentry{Quigg:2007dt}

\begin{figure}
    \includegraphics[width=8.5cm]{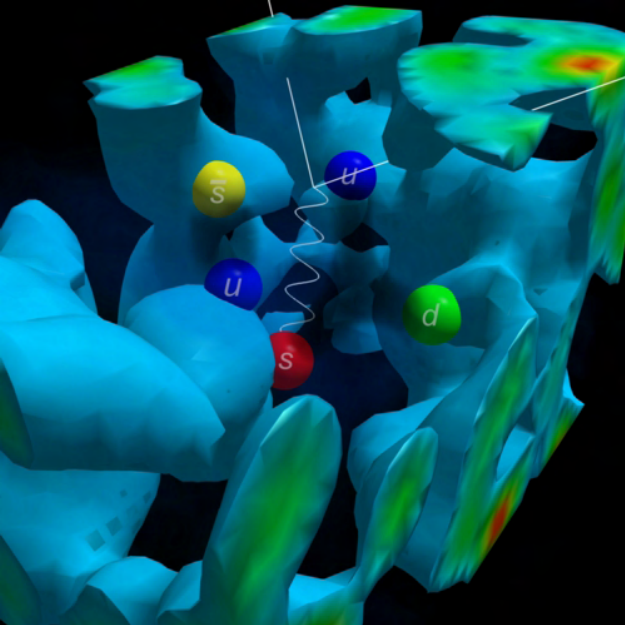}
	\caption{Artist's depiction of a proton (two $u$ and one $d$ quarks) 
        fluctuating into a $\Lambda$ baryon ($uds$) and a $K^+$ meson 
        ($\bar{s}u$), in the background of vacuum fluctations of gluons.
        The blob's false color scale denotes the intensity of the gluon 
        field at a given moment,
        proceeding through the rainbow from blue to red (most intense).
        From Derek Leinweber, CSSM, University of Adelaide
        (Ref.~\onlinecite{visqcd}).}
	\label{fig:derek}
\end{figure}

The development of \textit{lattice gauge theory} has made possible a
quantitative understanding of how these phenomena emerge at the
low-energy scale associated with confinement.
\citentry{Wilson:1974sk}
\noindent The essential ideas are described in 
\citentry{Rebbi:1983ip}
\citentry{Weingarten:1996ig}
\noindent and how it all began is recalled in
\citentry{Wilson:2004de}
\noindent Visualizations of the QCD vacuum, the structure of the 
proton, and other insights from lattice QCD are presented and explained at
\citentry{visqcd}
\noindent An example is shown in Fig.~\ref{fig:derek}, depicting the process 
$p\leftrightarrow\Lambda K^+$ on a background of the gluonic ground state.

Lattice gauge theory is yielding a growing range of nonperturbative 
computations of hadron properties that are needed to interpret 
experiments and observations in particle physics, nuclear physics, and 
astrophysics:
\citentry{Kronfeld:2008zza}
\noindent By 2008 lattice-QCD calculations of the hadron masses had 
been carried out with an accuracy of a few per cent, as discussed in 
Sec.~\ref{sec:hadronmasses}.

For particle physics, an important structural feature of QCD is that it 
can be married successfully to theories of the weak interaction, even 
though quarks carry color, but leptons do not.
\citentry{Weinberg:1973un}

\subsection{Textbooks}
\label{sec:qcd:texts}

Many books treat quantum chromodynamics, in whole or in part, 
from a modern point of view.
Among books addressed to general readers and undergraduate students, 
see Ref.~\onlinecite{Wilczek08} and
\citentry{'tHooft:1992rg}
\citentry{Nambu:1985xx}
\citentry{Watson:2004yp}

Several excellent textbooks are addressed to graduate students 
and researchers:
\citentry{Ellis:1991qj}
\citentry{Dissertori:2003pj}
\citentry{Yndurain:2006ui}
\citentry{Greiner:2007}
\citentry{Muta:2009vi} 
\citentry{Ioffe:2010}

Among many fine field-theory textbooks, 
\citentry{Sterman:1994ce}
\noindent is particularly inclined toward QCD and the issue of
factorization.

The biannual Review of Particle Physics (Ref.~\onlinecite{Amsler:2008zzb}) 
contains several concise reviews of QCD and other topics in particle physics.
For a well-chosen collection of longer review articles providing an
encyclopedic treatment of QCD, see
\citentry{Shifman:2001ck}

\subsection{Antecedent physical theories}
\label{sec:qcd:parents}

The modern gauge theory is the synthesis of several ideas.  For 
completeness we provide some historical and review references here.

\subsubsection{Flavor symmetry and current algebra}
\label{sec:fsca}

The idea of flavor symmetries underlying hadron masses and decay 
amplitudes predates QCD:
\citentry{Ne'eman:1961cd}
\citentry{GellMann:1961ky}
\citentry{GellMann:1962xb}
\noindent Early papers are collected in
\citentry{GellMann:1964xy}

Both weak and electromagnetic interactions of the strongly interacting 
particles (hadrons) are described by currents.
The $\mathrm{SU(3)_{flavor}}$ classification symmetry relates 
properties of the weak and electromagnetic interactions of hadrons.
Gell-Mann proposed that the charges associated with weak-interaction 
currents could be identified with $\mathrm{SU(3)_{flavor}}$ symmetry 
operators:
\citentry{GellMann:1964tf}
\noindent The current-algebra hypothesis states that the time 
components of the vector and axial-vector matrix elements satisfy 
quark-model equal-time commutation relations.
Current algebra fixes the strength of the leptonic and hadronic parts 
of the weak current, and it proved immensely fruitful for interactions 
involving pseudoscalar mesons.
In QCD, an $\mathrm{SU(3)_{flavor}}$ symmetry appears in the limit that 
the quark masses can be neglected.

A useful early review of current algebra can be found in
\citentry{Bjorken:1968ej}
\noindent and an early reprint volume with explanatory text is
\citentry{ADCA}
\noindent A later, somewhat more mature assessment is
\citentry{TJGCA}

\subsubsection{The original quark model}

The notion of fractionally-charged quarks was introduced in
\citentry{Gell-Mann:1964nj}
\citentry{Zweig:1981pd}
\citentry{Zweig:1964jf}
\noindent Zweig used the term ``aces'' for quarks.
An early review of the quark model is in
\citentry{Gell-Mann:1972ph}
\noindent and helpful compilations of references on the quark model appear in
\citentry{Greenberg:1982zv}
\citentry{Gasiorowicz:1981jz}

A challenge to these ideas came from the non-observation of free, 
fractionally-charged particles.
The current limits are collected in Ref.~\onlinecite{Amsler:2008zzb}, 
and descriptions of the techniques may be found in
\citentry{Lyons:1984pw}
\citentry{Smith:1987me}
\citentry{Perl:2009}
\noindent With confinement in QCD, however, the search for isolatable 
fractional charges is a somewhat more subtle subject, perhaps 
explaining why searches for fractionally-charged particles have been to 
no avail. 

\subsubsection{Quarks with color}

A second challenge to the quark model lay in the spin-and-statistics
puzzle for the baryons.
If the baryon $J = \cfrac{1}{2}$ octet and $J = \cfrac{3}{2}$ decuplet
are taken to be composites of three quarks, all in relative $s$-waves,
then the wave functions of the decuplet states appear to be symmetric in
space$\times$spin$\times$isospin, in conflict with the Pauli exclusion
principle.
As explicit examples, consider the $\Omega^-$, formed of three
(presumably) identical strange quarks, $sss$, or the $\Delta^{++}$, an
isospin-$\cfrac{3}{2}$ state made of three up quarks, $uuu$.
To reconcile the successes of the quark model with the requirement that
fermion wave functions be antisymmetric, it is necessary to hypothesize
that each quark flavor comes in three distinguishable species, which we
label by the primary colors red, green, and blue.
Baryon wave functions may then be antisymmetrized in color.
For a review of the role of color in models of hadrons, see
\citentry{Greenberg:1977ph}
\noindent Further observational evidence in favor of the color-triplet 
quark model is marshaled in
\citentry{Bardeen:1972xk}
\noindent For a critical look at circumstances under which the number 
of colors can be determined in $\pi^{0}\rightarrow \gamma \gamma$ 
decay, see 
\citentry{Bar:2001qk}
\noindent The cross section for the reaction $e^+e^-\to\mathrm{hadrons}$ 
(\cf\ Sec.~\ref{sec:Rmumu}) provides independent evidence that quarks 
are color triplets.
As discussed above, color attains a deeper dynamical meaning in QCD.

\subsubsection{Partons}
\label{sec:qcd:partons}

Meanwhile, high-energy scattering experiments showed signs of nucleon
substructure in the SLAC-MIT experiments on deeply inelastic
electron-nucleon scattering.
The structure functions that describe the internal structure of the target 
nucleon as seen by a virtual-photon probe depend in principle on two 
kinematic variables: the energy $\nu = E - E^\prime$ lost by the scattered 
electron and the four-momentum transfer, $Q^2$. 
At large values of $\nu$ and $Q^2$ 
the structure functions depend, to good 
approximation, only on the single dimensionless variable, $x = Q^2/2M\nu$ 
(where $M$ is the nucleon mass), as anticipated by 
\citentry{Bjorken:1968dy}
\noindent \textit{Bjorken scaling} implies that the virtual photon 
scatters off pointlike constituents; otherwise large values of $Q^2$ would 
resolve the size of the constituents.
An early overview is
\citentry{Kendall:1971tz}
\noindent The first observations are reported in
\citentry{Bloom:1969kc}
\citentry{Breidenbach:1969kd}
\noindent The experiments and their interpretation in terms of the
parton model, which regards the nucleon as a collection of quasifree
charged scattering centers, 
are reviewed in the Nobel Lectures,
\citentry{Taylor:1991ew}
\citentry{Kendall:1991np}
\citentry{Friedman:1991nq} 
\noindent and in the narrative,
\citentry{Riordan:1987gw}

A theoretical framework called the ``parton model'' based on pointlike 
constituents of unknown properties was developed in
\citentry{Feynman:1973xc}
\citentry{Bjorken:1969ja}
\citentry{Close:1979bt}
\noindent Complementary experiments in high-energy neutrino beams soon 
sealed the identification of the charged partons as quarks, and pointed 
to the importance of neutral partons later identified as the gluons of QCD. 
\citentry{Eichten:1973cs}
\noindent See Sec.~\ref{sec:DIS} for references to works covering the 
recent experiments.

The parton-model interpretation of high-transverse-momentum scattering
in hadron collisions was pioneered by 
\citentry{Berman:1971xz}
\citentry{Bjorken:1973kd}
\citentry{Ellis:1973nb}
\noindent and implemented in practical terms in
\citentry{Field:1976ve}
\citentry{Field:1977fa}
\citentry{Feynman:1978dt}

\subsubsection{Gauge invariance and Yang-Mills theory}

The idea that a theory of the strong nuclear interactions could be
derived from a non-Abelian symmetry such as isospin dates to the 
work of
\citentry{Yang:1954ek}
\citentry{RonShaw}

The development of the notions of gauge invariance is detailed in
\citentry{Jackson:2001ia}
\noindent and many useful readings are compiled in
\citentry{Cheng:1988wf}
\noindent The concepts and consequences of local gauge invariance are 
recalled in
\citentry{Mills:1989wj}
\noindent and the history of gauge theories is explored in
\citentry{O'Raifeartaigh:1997ia}
\citentry{O'Raifeartaigh:2000vm}
\noindent In particular, Shaw's thesis is reprinted and discussed in 
Chapter~9 of Ref.~\onlinecite{O'Raifeartaigh:1997ia}.
For an assessment of a half-century's development of gauge
symmetry, see
\citentry{'tHooft:2005ji}

For early attempts to build a realistic gauge theory of the
strong interactions, see Ref.~\onlinecite{Ne'eman:1961cd} and
\citentry{Sakurai:1960ju}
\noindent The idea of a vector gluon theory may be found in 
\citentry{Nambu:1981pg}
\noindent and the path from currents to a gauge theory of the strong
interactions is laid out in
\citentry{Fritzsch:1972jv}

\section{Theoretical Tools}
\label{sec:tools}

Like many a realistic physical theory, Yang-Mills theories defy exact 
solution.
To gain a theoretical understanding and, hence, to see whether QCD 
mirrors nature, several lines of attack are necessary.
We describe some of the literature behind several theoretical tools, 
ordered by increasing complexity.
Readers concerned mainly with the physical consequences of QCD may wish
to pass first to Sec.~\ref{sec:expt}.

\subsection{Symmetries}

\subsubsection{Light quarks}
\label{lightquarks}

The spontaneous breaking of chiral symmetries was studied before the 
advent of QCD, to explain why the mass of the pions is so much smaller 
than that of the nucleons.
\citentry{Nambu:1960xd}
\noindent A prominent feature of spontaneously broken symmetries in 
quantum field theories is the appearance of a massless particle:
\citentry{Goldstone:1961eq}
\citentry{Goldstone:1962es}
\noindent The massless states are called Nambu-Goldstone particles.
When a small amount of explicit symmetry breaking arises, as with pions,
these states acquire a small mass and are called 
pseudo-Nambu--Goldstone particles.

An informative toy model in which the nucleon mass arises 
essentially as a self-energy in analogy with the appearance of the mass 
gap in superconductivity was presented in
\citentry{Nambu:1961tp}
\citentry{Nambu:1961fr}
\noindent This construction, three years before the invention of quarks, 
prefigured our current understanding of the masses of strongly 
interacting particles in quantum chromodynamics.
The pions arose as light nucleon-antinucleon bound states, following 
the introduction of a tiny ``bare'' nucleon mass and spontaneous 
chiral-symmetry breaking.

Spontaneous symmetry breaking is common in physics, and
parallels to condensed-matter physics are drawn in
\citentry{Nambu:2009zzz}

Meanwhile, QCD explains the origin of chiral symmetry via the smallness 
of the up-, down-, and strange-quark masses---recall the 
discussion following \Eqn{eq:chiral}.
The spontaneous breaking is driven by the formation of a condensate of 
the light quarks, measured by the vacuum expectation value
$\langle 0|\bar{q}q|0\rangle$.
For a careful calculation see
\citentry{Fukaya:2009fh}
\noindent yielding (in the $\overline{\textrm{MS}}$ scheme at 2\gev)
\begin{equation}
    \langle 0|\bar{q}q|0\rangle = \left[242\pm 4^{+19}_{-18}\mev
        \right]^3,
\end{equation}
where the first error stems from Monte Carlo statistics and 
the second encompasses systematic effects, such as extrapolation to 
vanishingly small up- and down-quark masses.

\subsubsection{Anomalous chiral symmetries}
\label{sec:anomaly}

Among the chiral symmetries of light quarks, the flavor-singlet 
symmetry is special, because a quantum-mechanical effect, called the 
anomaly, breaks the classical conservation law.
This effect implies that the $\eta'$, unlike the pions and kaons, 
should not be a pseudo-Nambu--Goldstone particle with small mass.
\citentry{Ambrosino:2009sc}
\noindent The details of how this arises are connected to the 
nontrivial vacuum structure of QCD (discussed in Sec.~\ref{sec:vacuum}): 
\citentry{'tHooft:1986nc}
\citentry{diVecchia:1980ve}
\citentry{Shifman:1991zk}

The phase of the quark mass matrix combines with the coefficient of
$\varepsilon_{\mu\nu\rho\sigma}\tr(G^{\mu\nu}G^{\rho\sigma})$
in the Lagrangian to cause effects that violate CP symmetry.
Curiously, this combination---the difference of two quantities with 
starkly distinct origins---is constrained by the neutron electric 
dipole moment to be $\sim10^{-11}$.
The \emph{strong CP problem} was clearly posed, and a still-popular 
resolution proposed in
\citentry{Peccei:1977hh}
\noindent Further possible resolutions are explained in
\citentry{Dine:2000cj}
\noindent The Peccei-Quinn solution requires a new particle, the axion, 
with several implications for particle physics and, possibly, cosmology.
These connections, and the status of axion searches, are reviewed in
\citentry{Kim:2008hd}

\subsubsection{Heavy quarks}

Hadrons containing heavy quarks exhibit simplifying features.
In a bound state with one heavy quark, and any number of light quarks 
and gluons, the identity (flavor or spin) of the heavy quark alters the
dynamics very little, because the heavy quark sits essentially at rest 
inside the hadron:
\citentry{Shuryak:1981fza}
\citentry{Shifman:1986sm}
\noindent The center of mass of the hadron and the heavy quark are 
essentially the same, with the light degrees of freedom in orbit around 
the heavy quark.
A set of approximate symmetries emerge, the heavy-quark flavor and spin 
symmetries.
\citentry{Isgur:1989vq}
\citentry{Isgur:1990ed}

In a meson with a heavy quark and corresponding antiquark, the two orbit 
each other.
The velocity depends on the heavy-quark mass, but the spin decouples (to 
leading order), in analogy with QED applied to atomic physics.
\citentry{Caswell:1985ui}

\subsection{Potential models}
\label{sec:QQ}

The observation that asymptotic freedom suggests nonrelativistic atoms
of heavy quarks and antiquarks is due to 
\citentry{Appelquist:1975zd}
\noindent The nonrelativistic description was elaborated in
\citentry{Eichten:1978tg}
\citentry{Eichten:1979ms}
\noindent A midterm review of potential models can be found in
\citentry{Kwong:1987mj}
\noindent and newer reviews include
\citentry{Eichten:2007qx}
\noindent More recently this line of research has been addressed 
further through effective field theories (see Sec.~\ref{sec:QQEFT}).

\subsection{Renormalization and factorization}
\label{sec:rnf}

The renormalization group as a technique for summing to all orders in
perturbation theory in electrodynamics was invented by
\citentry{Stueckelberg:1953dz}
\citentry{Gell-Mann:1954fq}
\noindent A clear statement of the algorithm and a thorough review of
early applications appears in 
\citentry{Bogolyubov:1980nc}
\noindent The modern formulation of the renormalization group equations
is due to
\citentry{Callan:1970yg}
\citentry{Symanzik:1970rt}
\noindent The power of renormalization group methods for a wide range
of physical problems was recognized by
\citentry{Wilson:1970ag}
\citentry{Wilson:1979qg}
\noindent A fascinating survey with many references is
\citentry{Wilson:1993dy}

The theoretical apparatus required for a general analysis of
quantum corrections and their implications for a running coupling
constant is presented in
\citentry{Coleman:1974ae}
\citentry{Balian:1976vq} 
\citentry{Peterman:1978tb}
\citentry{Collins:1984xc}

The ability to predict characteristics of high-energy reactions rests
on parton-hadron duality and on separating short-distance
hard-scattering matrix elements described by perturbative QCD from
long-distance (nonperturbative) effects related to hadronic structure.
\textit{Duality} refers to the observation that inclusive hadronic
observables may be computed in terms of quark and gluon degrees of
freedom.
These ideas are reviewed and confronted with recent experimental data in
\citentry{Melnitchouk:2005zr}
\noindent updating the classic reference:
\citentry{Azimov:1984np}
\noindent The distinction between short-distance and long-distance (or 
short and long time scales) is reminiscent of the Born-Oppenheimer 
approximation in molecular physics.
The \textit{factorization} of amplitudes and cross sections into parton 
distribution functions, elementary scattering amplitudes, and 
fragmentation functions that describe how partons materialize into 
hadrons, was an element of the exploratory studies reported in 
Ref.~\onlinecite{Field:1976ve}. 
Within the framework of QCD, factorization has been proved in many 
settings:
\citentry{Collins:1987pm} 
\citentry{Collins:1989gx}
\citentry{Ellis:1978ty}

The short-distance behavior of quantum field theories, including QCD, 
is clarified by the operator-product expansion:
\citentry{Wilson:1972ee}
\noindent in which a product of operators is related to a series of 
local operators.

Factorization of amplitudes hinges on an understanding of universal 
behaviors of field theory when massless particles become soft or if two 
massless particles become collinear.
These features already appear in QED for the scattering of a 
high-energy photon off an electron:
\citentry{Sudakov:1954sw}
\noindent and were generalized to quarks and QCD in
\citentry{Sen:1981sd}
\citentry{Magnea:1990zb}
\noindent For processes with four or more external particles, as in 
collisions, these aspects were further developed in
\citentry{Catani:1998bh}
\citentry{Kidonakis:1998nf}
\citentry{Sterman:2002qn}
\noindent Understanding the singularity structure of massless 
gauge-theory amplitudes continues to be germane, with application to 
high-energy collider physics:
\citentry{Aybat:2006wq}
\citentry{Gardi:2009qi}
\citentry{Dixon:2009ur}
\noindent For similar results derived using effective-field-theory 
techniques, see Sec.~\ref{sec:scet}.

Techniques of factorization have been extended to exclusive processes 
in 
\citentry{Lepage:1980fj}
\noindent and adapted to decays of hadrons containing a heavy quark in
\citentry{Beneke:2000ry}

The study of higher orders in perturbation theory, particularly the 
renormalization parts, can anticipate the pattern of 
nonperturbative effects.
A standard review is
\citentry{Beneke:1998ui}
\noindent An intriguing feature of these effects makes the definition 
of quark masses somewhat subtle.
The so-called ``pole mass,'' which corresponds closely to the classical 
notion of mass, is well-defined in perturbation theory:
\citentry{Kronfeld:1998di}
\noindent yet the perturbative series signals the necessity of 
nonperturbative effects:
\citentry{Beneke:1994sw}
\citentry{Bigi:1994em}
\noindent As a consequence, quark masses reported below are 
renormalized Lagrangian masses.

\subsection{Unitarity and analyticity}

Underlying the notion of parton-hadron duality, which enters into many 
applications of factorization, are unitarity and analyticity.
Unitarity means merely that quantum mechanics (and, hence, quantum 
field theory) preserves probability, 
thereby imposing limits on scattering amplitudes and related quantities.
Analyticity means that scattering amplitudes are 
analytic functions of kinematic variables, apart from poles or branch 
cuts, which correspond to stable particles and resonances or 
multi-particle thresholds, respectively.

These ideas and the formalism of quantum field theory can be used to 
derive semi-quantitative and, sometimes, quantitative dynamical 
information.
This approach goes under the name ``QCD sum rules'' and started with
\citentry{Shifman:1978bx}
\citentry{Shifman:1978by}
\noindent An early, well-regarded review is
\citentry{Reinders:1984sr}
\noindent A review and reprint volume is
\citentry{Shifman:1992xu}
\noindent A more recent monograph explaining QCD sum rules is
\citentry{Narison:2002pw}
\noindent and several pedagogical reviews of applications can be 
found in Ref.~\onlinecite{Shifman:2001ck}.

\subsection{Effective field theories}

Effective field theories isolate important low-energy degrees of 
freedom, absorbing the effects of highly virtual processes, such as 
those of high-mass particles, into coupling strengths of interactions.
For capsule reviews, see
\citentry{Georgi:1994qn}
\citentry{Ecker:2005ny}
\citentry{Burgess:2007pt}
\noindent Two classes of effective field theories are employed to study 
QCD, one in which (light) quarks and gluons remain the basic degrees of 
freedom, and another treating hadrons as fundamental.
In both cases, the power of the method is to retain and respect 
symmetry, renormalization, unitarity, analyticity, and cluster 
decomposition.

\subsubsection{Chiral perturbation theory}

The consequences of spontaneously broken symmetries are encoded in current 
algebra (see Sec.~\ref{sec:fsca}) and can be summarized in an effective 
Lagrangian for pions:
\citentry{Weinberg:1966fm}
\citentry{Weinberg:1968de}
\noindent The formalism was extended to general patterns of spontaneous 
symmetry breaking in
\citentry{Coleman:1969sm}
\citentry{Callan:1969sn}

For hadron dynamics chiral Lagrangians were developed further in
\citentry{Dashen:1969eg}
\citentry{Dashen:1969ez}
\citentry{Li:1971vr}
\noindent An early review is
\citentry{Pagels:1974se}

The connection with the quark model is developed in
\citentry{Manohar:1983md}

Chiral Lagrangians were then exploited to develop a systematic 
low-energy expansion, called chiral perturbation theory ($\chi$PT):
\citentry{Weinberg:1978kz}
\citentry{Gasser:1983yg}
\noindent An excellent place to start learning the modern perspective is
\citentry{Leutwyler:1993iq}

This is now a subject with broad applications, describing, for example, 
the pion and kaon clouds surrounding a nucleon.
This material is pedagogically reviewed in
\citentry{Scherer:2002tk}
\citentry{Scherer:2009bt}

States with nucleonic properties can also arise from soliton 
configurations of the pion field, which was first noticed before the 
advent of QCD:
\citentry{Skyrme:1959vp}
\noindent The so-called Skyrmion approach to the nucleon enjoyed a 
renaissance in the 1980s, reviewed in
\citentry{Zahed:1986qz}

\subsubsection{Heavy-quark effective theory and nonrelativistic~QCD}
\label{sec:QQEFT}

The simpler dynamics of heavy-quark systems lend themselves to effective 
field theories.
For heavy-light hadrons (those with one heavy quark), this insight led 
to the development of the heavy-quark effective theory (HQET) in
\citentry{Eichten:1987xu}
\citentry{Eichten:1989zv}
\citentry{Eichten:1990vp}
\citentry{Georgi:1990um}
\citentry{Grinstein:1990mj}
\noindent Some pedagogical reviews are
\citentry{Grinstein:1992ss}
\citentry{Neubert:1994mb}
\citentry{Bigi:1997fj}
\noindent and a textbook is
\citentry{Manohar:2000dt}

In quarkonium, a heavy quark's velocity is larger than in a heavy-light 
hadron.
The appropriate effective field theory has the same Lagrangian as HQET, 
but the relative importance of various interactions is different.
This field theory is called nonrelativistic QCD (NRQCD) and was first 
developed for bound-state problems in Ref.~\onlinecite{Caswell:1985ui}
and
\citentry{Lepage:1987gg}
\citentry{Thacker:1990bm}
\noindent The classification of NRQCD interactions, focusing on the quarkonium 
spectrum, was further elucidated in
\citentry{Lepage:1992tx}

NRQCD was extended to encompass decay, production, and annihilation in
\citentry{Bodwin:1992ye}
\citentry{Bodwin:1994jh}

In some applications, the QCD coupling $\alphas$ is small 
at both the heavy-quark mass and heavy-quark momentum scales:
\citentry{Beneke:1997jm}
\noindent Then the appropriate effective field theory is potential 
NRQCD (PNRQCD): 
\citentry{Brambilla:1999xf}
\noindent PNRQCD provides a field-theoretic basis for understanding the 
success of the potential models of Sec.~\ref{sec:QQ}.
For a review, consult
\citentry{Brambilla:2004jw}

NRQCD and PNRQCD have also been used to understand top-quark pair 
production at threshold.
Top quarks decay before toponium forms:
\citentry{Bigi:1986jk}
\citentry{Fadin:1987wz}
\citentry{Fadin:1988fn}
\noindent but top and antitop still orbit each other during their 
fleeting existence.
A useful review is
\citentry{Hoang:2000yr}

\subsubsection{Soft collinear effective theory}
\label{sec:scet}

In high-energy amplitudes, one often considers a jet of particles, the 
details of which are not detected.
The semi-inclusive nature of jets circumvents issues of infrared and 
collinear divergences, much like the Bloch-Nordsieck mechanism in QED:
\citentry{Bloch:1937pw}
\citentry{Kinoshita:1962ur}
\citentry{Lee:1964is}
\noindent The infrared and collinear degrees of freedom can be isolated 
in the soft collinear effective theory (SCET), first established for 
decays of $B$~mesons:
\citentry{Bauer:2000ew}
\citentry{Bauer:2000yr}
\citentry{Bauer:2001yt}
\citentry{Beneke:2002ph}
\noindent Meanwhile, SCET has been applied to many high-energy scattering 
processes, starting with
\citentry{Bauer:2002nz}
\noindent and more recently to many aspects of jets:
\citentry{Becher:2009qa}
\citentry{Beneke:2009rj}
\citentry{Chiu:2009mg}
\citentry{Stewart:2009yx}
\citentry{Mantry:2009qz}
\citentry{Ellis:2009wj}

\vspace*{-1em}
\subsection{Lattice gauge theory}

With an explicit definition of its ultraviolet behavior,
lattice gauge theory lends itself to computational methods, 
essentially integrating the functional integral of QCD numerically:
\citentry{Wilson:1979wp}
\noindent The first study connecting the confining regime to asymptotic 
freedom appeared in
\citentry{Creutz:1980zw}
\citentry{Creutz:1980wj}
\noindent A useful reprint collection of early work is
\citentry{Rebbi:1984tx}

There are several good textbooks on lattice gauge theory, including
\citentry{Creutz:1984mg}
\citentry{Montvay:1994cy}
\citentry{Smit:2002ug}
\citentry{Rothe:2005nw}
\citentry{DeGrand:2006zz}
\citentry{Gattringer:2010zz}

Lattice gauge theory is also the foundation of attempts at rigorous 
construction of gauge theories:
\citentry{Seiler:1982pw}
\citentry{Glimm:1987ng}

For many years, numerical lattice-QCD calculations omitted the computationally 
very demanding contribution of sea quarks (quark-antiquark pairs that 
fluctuate out of the vacuum), leading to uncontrolled uncertainties.
The first demonstration that incorporation of sea-quark effects brings a 
wide variety of computed hadron properties into agreement with 
experiment is
\citentry{Davies:2003ik}
\noindent The maturation of numerical lattice QCD is discussed in
\citentry{DeTar:2004tn}
\noindent With these developments it is now possible to compute the 
hadron masses with a few percent precision:
\citentry{Kronfeld:2008zz}
\citentry{Wilczek:2008zz}
\noindent and make predictions of hadronic properties needed to interpret 
experiments:
\citentry{Kronfeld:2006sk}
\noindent A more detailed comparison of lattice-QCD calculations with 
experiment is given in Sec.~\ref{sec:expt}.

Numerical lattice QCD is not merely a brute-force approach, but a 
synthesis of computation and effective field theories.
Errors from nonzero lattice spacing are controlled with Symanzik's 
effective theory of cutoff effects:
\citentry{Symanzik:1983dc}
\citentry{Symanzik:1983gh}
\noindent work that grew out of Ref.~\onlinecite{Symanzik:1970rt}.
Errors from finite volume can be controlled with general properties of 
massive field theories on a torus:
\citentry{Luscher:1985dn}
\citentry{Gasser:1987zq}
\citentry{Luscher:1990ux}
\noindent The light quarks in computer simulations often have masses 
larger than those of the up and down quarks, but the extrapolation in 
quark mass can be guided by adapting chiral perturbation theory:
\citentry{Bar:2004xp}
\citentry{Bernard:2006gx}
\noindent The charmed and bottom quarks often have masses close to the 
ultraviolet cutoff (introduced by the lattice), but the effects can be 
understood with HQET and NRQCD:
\citentry{Kronfeld:2003sd}
\noindent The idea that lattice QCD is a synthesis of computational 
and theoretical physics is explored in
\citentry{Kronfeld:2002pi}

Lattice gauge theory and chiral symmetry coexist uneasily:
\citentry{Nielsen:1981hk}
\noindent The efforts to understand and overcome these difficulties, 
for theories like QCD, is reviewed in
\citentry{Neuberger:2001nb}

Lattice gauge theory, with its rigorous mathematical definition, is a 
suitable arena for deriving mass inequalities:
\citentry{Weingarten:1983uj}
\citentry{Witten:1983ut}
\noindent These and related developments have been reviewed in
\citentry{Nussinov:1999sx}

\subsection{The QCD vacuum and confinement}
\label{sec:vacuum}

The space of all non-Abelian gauge fields is not simply connected, but 
consists of sectors labeled by an integer~$n$.
The sectors arise when trying to satisfy a \emph{gauge condition}, 
namely to specify $\omega(x)$ in order to choose 
one representative field $B_\mu(x)$ among all those related by 
Eq.~(\ref{eq:gt:B}).

In some cases it is necessary to specify different conditions in different 
regions of spacetime, and then it turns out that $\omega(x)$ on the 
overlaps of the regions is an $n$-to-one mapping 
onto~$\mathrm{SU}(N_{\mathrm{c}})$.
In the quantum theory, tunneling can occur between the different 
sectors, and the tunneling events are called ``instantons.''
A classic discussion can be found in
\citentry{Coleman:1978ae}
\noindent Some further features appear at nonzero temperature:
\citentry{Gross:1980br}

Because of these sectors, the QCD Lagrangian, Eq.~(\ref{eq:qcdlag}), 
can contain a term proportional to 
$\varepsilon_{\mu\nu\rho\sigma}\tr(G^{\mu\nu}G^{\rho\sigma})$.
The physical implication of this term is a possible violation of CP 
symmetry, as is discussed further in Sec.~\ref{sec:anomaly}.

It is widely believed that the nontrivial vacuum structure is connected 
to the special features of QCD, notably confinement.
Opinion is divided whether instantons, i.e., the tunneling events, play 
the principal role, or whether strong quantum fluctuations do.
The case for instantons can be traced from
\citentry{Shuryak:1996wx}
\citentry{Shuryak:1997vd}
\noindent and the case for fluctuations from
\citentry{Witten:1978bc}
\citentry{Horvath:2001ir}

Another approach to confinement starts with the observation that any 
gauge condition has more than one solution:
\citentry{Gribov:1977wm}
In the Coulomb gauge one demands $\bm{\nabla}\cdot\bm{A}=0$;
further demanding a unique resolution of the Gribov ambiguity, one finds, 
with some assumptions, a confining potential:
\citentry{Zwanziger:1998ez}

A string picture of confinement emerges naturally from the perturbative 
properties of QCD.
The energy required to separate a quark and antiquark,
\begin{equation}
    E = \sigma R,
    \label{eq:dualstringtension}
\end{equation}
is proportional to the string tension $\sigma$ and the separation $R$.
Furthermore, the property of asymptotic freedom means that the 
``dielectric constant'' of the QCD vacuum is $\varepsilon_{\mathrm{QCD}}<1$,
in contrast to the familiar result for a dielectric substance, $\varepsilon>1$.
The QCD vacuum is thus a \textit{dia-electric} medium.
An electrostatic analogy leads to a heuristic understanding of confinement.
It is energetically favorable for a test charge placed in a very 
effective dia-electric medium to carve out a bubble in which 
$\varepsilon = 1$.
In the limit of a perfect dia-electric medium, the bubble radius and 
the energy stored in the electric field tend to infinity.
In contrast, the radius of the bubble surrounding a test dipole placed in 
the medium occupies a finite volume, even in the perfect dia-electric 
limit, because the field lines need not extend to infinity.
\citentry{Kogut:1974sn}
\noindent The dia-electric analogy is reviewed in Sec.~8.8 of
\citentry{GTSWEMI}
\noindent The physical picture is highly similar to MIT bag model:
\citentry{Chodos:1974je}

The exclusion of chromoelectric flux from the QCD vacuum is reminiscent 
of the exclusion of magnetic flux from a type-II superconductor.
In a dual version of the Meissner effect, with the roles of electric 
and magnetic properties swapped, 
the chromoelectric field between a separating quark and 
antiquark takes the form of an Abrikosov flux tube.
For an introduction and tests of the picture, see   
\citentry{DiGiacomo:1999fa}

\begin{figure}[bt] 
	\includegraphics[width=8.5cm]{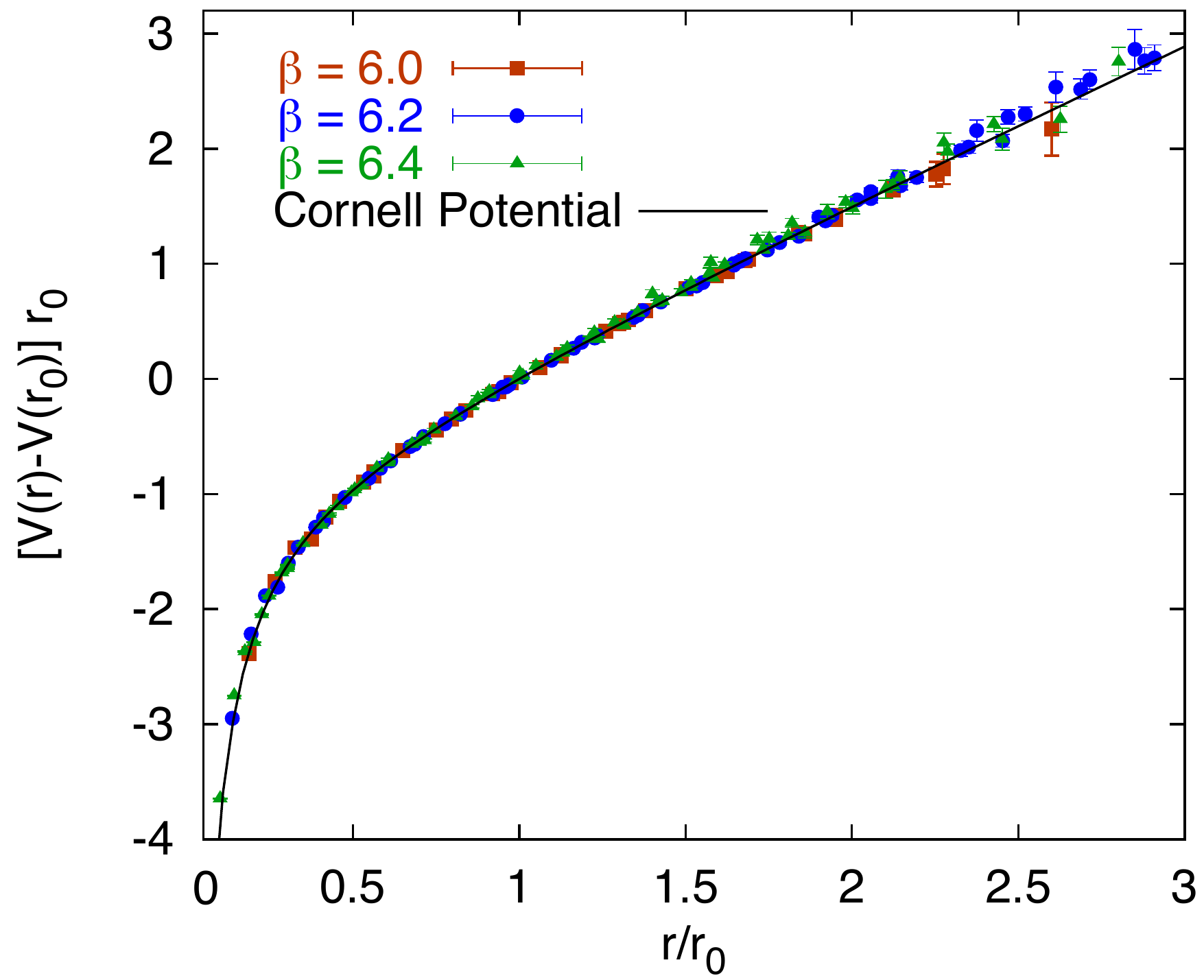}
	\caption{The potential energy $V(r)$ between static sources of 
    color (in an approximation without sea quarks).
    The zero of energy and the units are set by a conventional 
    distance~$r_0$, defined by $r_0^2\,dV/dr=1.65$.
    The data points are from lattice QCD, generated at several values 
    of $\beta=6/g^2$, which---via dimensional transmutation---corresponds 
    to varying the spacing between lattice sites.
    The black curve is a fit of these data to the potential model of 
    Refs.~\onlinecite{Eichten:1978tg,Eichten:1979ms}.
    From Ref.~\onlinecite{Bali:2000gf}.}
	\label{fig:potential}
\end{figure}

Lattice gauge theory (Ref.~\onlinecite{Wilson:1974sk}) was originally 
invented to understand confinement.
Reviews of more recent analytical and numerical work can be found in
\citentry{Greensite:2003bk}
\citentry{Alkofer:2006fu}
\noindent The connection between QCD potentials, spectroscopy, and 
confinement is reviewed in
\citentry{Bali:2000gf}
An important theme in Ref.~\onlinecite{Bali:2000gf} is the 
lattice-QCD computation of the potential energy between static sources 
of color.
As shown in Fig.~\ref{fig:potential}, the potential looks Coulombic at 
short distances, in accord with asymptotic freedom, and linear at long 
distances, in accord with Eq.~(\ref{eq:dualstringtension}).

A series of conferences is devoted to the confinement problem.
Their agendas and proceedings can be traced from 
\citentry{qchs9}

\subsection{Dyson-Schwinger Equations}
\label{sec:DSE}

A fruitful continuum approach to nonperturbative dynamics is based on 
the infinite tower of Dyson-Schwinger equations, coupled integral 
equations that relate the Green functions of a field theory to each 
other.
Solving these equations provides a solution of the theory, in that a 
field theory is completely defined by all of its $n$-point Green 
functions.
A good starting point is     
\citentry{Roberts:1994dr}
\noindent A newer review, focused on mesons, is
\citentry{Maris:2003vk}
\noindent and a broader survey of results from this approach can be 
found in
\citentry{Chang:2010jq}
\noindent Truncations and approximations of Dyson-Schwinger equations 
can shed light on confinement:
\citentry{vonSmekal:1997vx}
\noindent Finally, these techniques have been extended to QCD 
thermodynamics in
\citentry{Hong:1999fh}
\citentry{Roberts:2000aa}

\subsection{Perturbative amplitudes}

A key consequence of factorization is to relate amplitudes for (some) 
hadronic processes to underlying processes of quarks and gluons.
Parton amplitudes can be computed via Feynman diagrams, as discussed in 
Refs.~\onlinecite{Ellis:1991qj,Yndurain:2006ui,Greiner:2007,%
Muta:2009vi,Ioffe:2010,Sterman:1994ce}.
As the complexity of the process increases, however, this approach 
becomes intractable.
Remarkably, QCD amplitudes are simpler than the individual diagrams 
might suggest:
\citentry{Parke:1986gb}
\noindent Perturbative QCD amplitudes also are related by recursion in 
the number of scattered gluons:
\citentry{Berends:1987me}
\noindent For an older review that remains useful for graduate 
students, see
\citentry{Mangano:1990by}

The simplifications can be related to deep connections between 
Yang-Mills theories and string theories:
\citentry{Bern:1991aq}
\citentry{Witten:2003nn}
\noindent A parallel, and perhaps, even more fruitful, alternative to 
Feynman diagrams starts with constraints of unitarity:
\citentry{Bern:1994zx}

The first decade of the 2000s witnessed rapid conceptual and technical
development of these two sets of ideas, by many researchers, too many 
to list here.
The review
\citentry{Bern:2007dw}
\noindent contains a comprehensive set of references, and the most 
recent developments are discussed in 
\citentry{Berger:2009zb}

\subsection{Parton-shower Monte Carlo programs}

In a high-energy collision, although the parton-scattering can be 
factorized and computed in perturbation theory, a description of the 
full event is complicated first by radiation of gluons and $q\bar{q}$ 
pairs and later by the formation of hadrons.
Several computer codes have been developed to automate the calculation 
of the initial parton scatter, treat the shower of partons, and model 
the hadronization.
Useful reviews to the concepts can be found in
\citentry{Mangano:2005dj}
\citentry{Sjostrand:2009ad}
\noindent and a hands-on guide is
\citentry{Dobbs:2004qw}

\subsection{Extensions of QCD}

QCD belongs to a class of Yang-Mills theories, and further information 
can be gleaned by varying the number of colors, $N_{\rm c}$, and the 
number of flavors, $n_{\rm f}$.
Some classic and useful references on QCD as $N_{\rm c}\to\infty$ are
\citentry{'tHooft:1973jz}
\citentry{Witten:1979kh}
\citentry{Das:1987nb}
\citentry{Cohen:1995ch}
\citentry{Jenkins:1998wy}

Supersymmetry is a spacetime symmetry connecting bosonic and fermionic 
representations of the Poincar\'e group.
Gauge theories with supersymmetry enjoy some simplifying features:
\citentry{Seiberg:1994rs}
\citentry{Witten:1998zw}
\noindent leading to interesting relations between strongly-coupled 
gauge theories of certain $(N_{\rm c},n_{\rm f})$ and weakly-coupled 
dual gauge theories with $(N'_{\rm c},n'_{\rm f})$.

\subsection{String theory}

String theory is a mathematical description of particles as vibrational 
modes of one-dimensional objects, instead of as points.
First developed as a model of hadrons, string theory fell out of favor
after the rise of QCD.
But it has enjoyed a tremendous interest as a unifying theory of quantum 
mechanics and gravity, spurring a vast literature in mathematical 
physics.
Now string theory has come full circle, with string techniques applied 
to gauge theories, starting with
\citentry{Maldacena:1997re}
\noindent An excellent pedagogical introduction is given in
\citentry{Klebanov:2000me}

Several developments address hadron properties, for example
\citentry{Csaki:1998qr}
\citentry{Brower:2000rp}
\citentry{Polchinski:2000uf}
\citentry{Polchinski:2001tt}
\citentry{Brower:2006ea}

\section{Confronting QCD with Experiment}
\label{sec:expt}

\subsection{\boldmath Running of $\alphas$}
\label{sec:runalpharun}

A fundamental consequence of QCD is the property of asymptotic freedom, 
the decrease of the strong coupling constant $\alphas(Q)$ with 
increasing values of the momentum scale, $Q$. In first approximation 
(see Sec.~\ref{sec:qcd:first}), we expect a linear increase of 
$1/\alphas(Q)$ with $\log{Q}$:
$$  
	\frac{1}{\alphas(Q)} = \frac{1}{\alphas(\mu)} + 
		\frac{33-2n_{\mathrm{f}}}{6\pi}
		\log\left(\frac{Q}{\mu}\right),
    \reqno{eq:asevol}
$$
so long as the number of active quark flavors, $n_{\mathrm{f}}$, does 
not exceed 16.
In fact, the scale dependence of $\alphas$ to be expected 
in QCD has been computed to order~$\alphas^5$:
\citentry{vanRitbergen:1997va}
\noindent The decrease of $\alphas$ with $Q$ has been demonstrated by
measurements in many experimental settings 
(Ref.~\onlinecite{Amsler:2008zzb}).
Over the past decade, the precision of $\alphas$ 
determinations has improved dramatically, thanks to a plethora of 
results from various processes aided by improved calculations at higher 
orders in perturbation theory.
The progress is reviewed, and critically evaluated, in
\citentry{Bethke:2009jm}
\noindent which draws particular attention to the high level of recent 
activity in the area of hadronic $\tau$ decays.

\begin{figure}[bt] 
	\includegraphics[width=8.5cm]{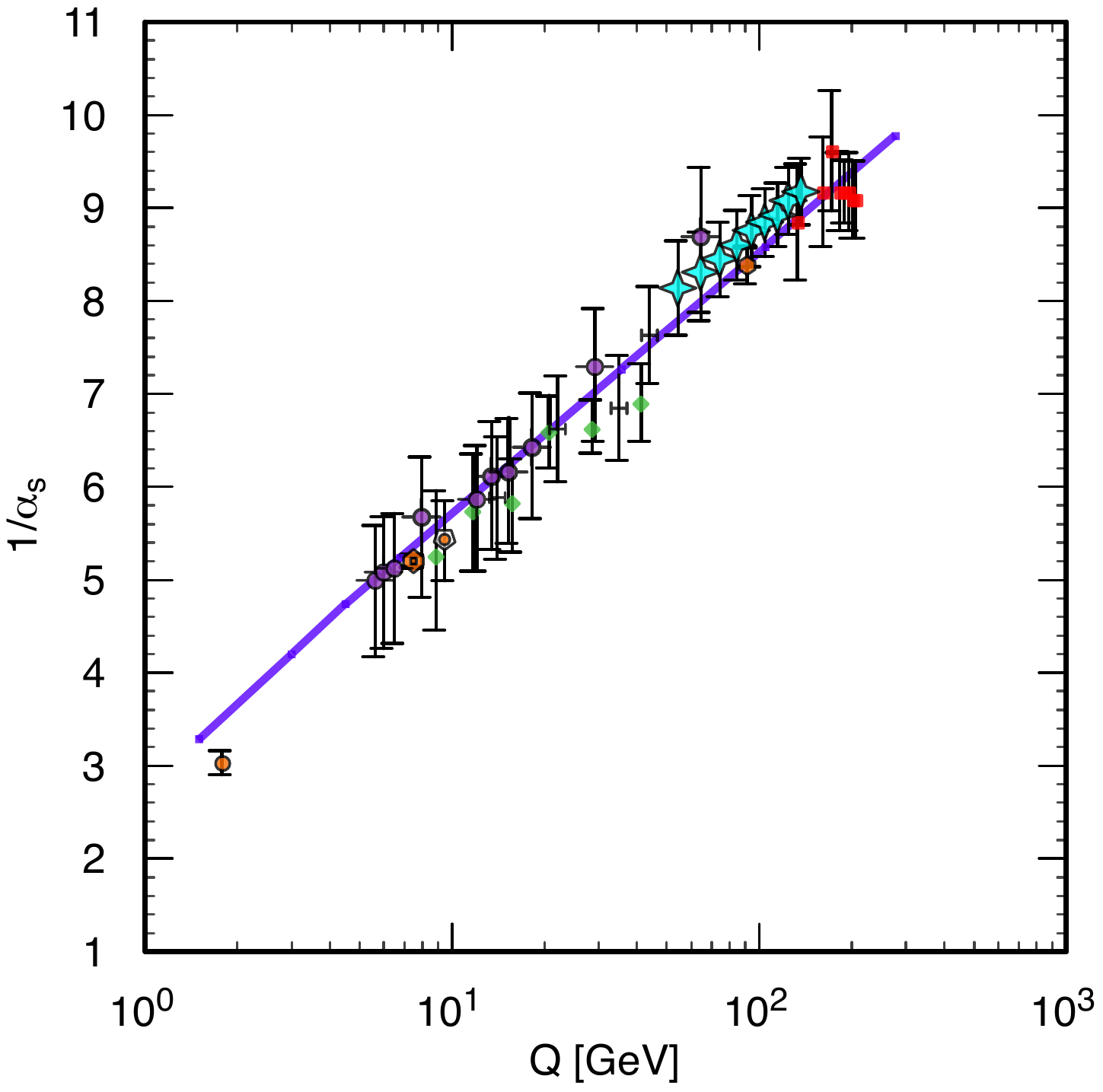}
	\caption{Measurements of the strong coupling $1/\alphas(Q)$ as a
		function of the energy scale $\ln{Q}$.
        In addition to hadronic $\tau$-decay, quarkonium, $\Upsilon$ 
        decay, and $Z^0$-pole values from Ref.~\onlinecite{Bethke:2009jm}, 
        we display black crosses: $e^+e^-$ collisions 
        (Ref.~\onlinecite{MovillaFernandez:2002zz});
        red squares: $e^+e^-$ collisions (Ref.~\onlinecite{Achard:2002kv}); 
        green diamonds: $e^\pm p$ collisions (Ref.~\onlinecite{Chekanov:2006yc});
        barred purple circles: $e^\pm p$ collisions
        (Refs.~\onlinecite{Aaron:2009vs,Collaboration:2009he}); 
        cyan crosses: $\bar{p}p$ collisions (Ref.~\onlinecite{Abazov:2009nc});
		average value of $\alphas(M_Z)$, in 4-loop approximation and
		using 3-loop threshold matching at the heavy-quark pole masses
		$m_c = 1.5\gev$ and $m_b = 4.7\gev$.}
	\label{fig:alpharun}
\end{figure}

A representative selection of experimental determinations is shown, 
together with the evolution expected in QCD, in Fig.~\ref{fig:alpharun}.
We have drawn the displayed values from Ref.~\onlinecite{Bethke:2009jm},
together with determinations from $e^+e^-$ event shapes reported in
\citentry{MovillaFernandez:2002zz}
\citentry{Achard:2002kv}
\noindent from jet studies in $e^{\pm}p$ scattering reported in
\citentry{Chekanov:2006yc}
\citentry{Aaron:2009vs}
\citentry{Collaboration:2009he}
\noindent and the running coupling constant inferred from inclusive jet
production in $\bar{p}p$ collisions,
\citentry{Abazov:2009nc}
\noindent The trend toward asymptotic freedom is clear, and the 
agreement with the predicted evolution is excellent, within the 
uncertainties in the measurements.
An interesting challenge for the future will be to 
measure~$\alphas(Q)$ with precision sufficient to detect 
the expected change of slope at the top-quark threshold.

It is conventional, and enlightening, to rewrite the evolution equation
(\ref{eq:asevol}) in the form
\begin{equation}
	\frac{1}{\alphas(Q)} = \frac{33 - 2n_{\mathrm{f}}}{6\pi}
		\log\left(\frac{Q}{\Lambda_{\mathrm{QCD}}}\right),
	\label{eq:lambdadef}
\end{equation}
where $\Lambda_{\mathrm{QCD}}$ is the QCD scale parameter, with
dimensions of energy.
(A generalization beyond leading order is given in Sec.~9 of
Ref.~\onlinecite{Amsler:2008zzb}.)
Several subtleties attend this simple and useful parametrization.
First, if we enforce the requirement that $\alphas(Q)$ be 
continuous at flavor thresholds, then $\Lambda_{\mathrm{QCD}}$ must 
depend on the number of active quark flavors.
Second, the value of $\Lambda_{\mathrm{QCD}}$ depends on the
renormalization scheme; the canonical choice is the modified minimal
subtraction ($\overline{\mathrm{MS}}$) scheme introduced in
\citentry{Bardeen:1978yd}
\noindent The $n_{\mathrm{f}}$ and scheme dependence is given via 
labels on $\Lambda$, \eg, 
$\Lambda^{(n_{\mathrm{f}})}_{\overline{\mathrm{MS}}}$.
Representative estimates of the QCD scale are
$\Lambda^{\mathrm{(5)}}_{\overline{\mathrm{MS}}} = 213\mev$,
$\Lambda^{\mathrm{(4)}}_{\overline{\mathrm{MS}}} = 296\mev$, and
$\Lambda^{\mathrm{(3)}}_{\overline{\mathrm{MS}}} = 338\mev$
(Ref.~\onlinecite{Bethke:2009jm}).
The appearance of a dimensional quantity to parametrize the running 
coupling is sometimes called ``dimensional transmutation.''

The theoretical underpinnings of high-precision determinations of 
$\alphas$ in deeply inelastic scattering are presented in 
\citentry{Blumlein:2006be}
\noindent which provides a detailed analysis of remaining uncertainties 
in the QCD scale.

For another critical assessment of the theoretical analyses that underlie
determinations of $\alphas$, see
\citentry{Prosperi:2006hx}

\begin{figure}[tb] 
	\includegraphics[width=8.0cm]{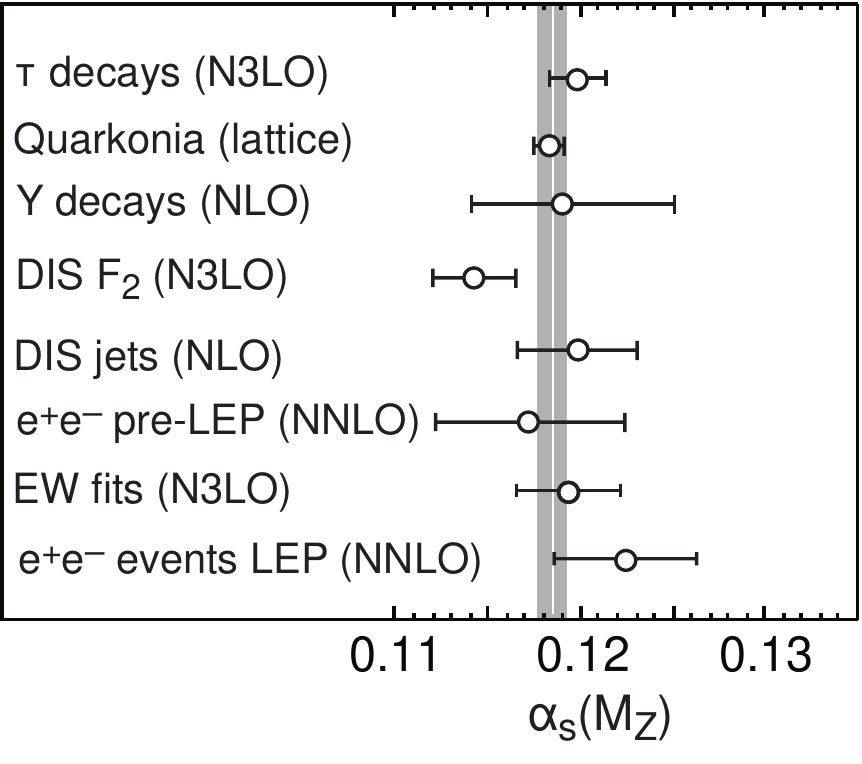}
    \caption{Determinations of $\alphas(M_{Z})$ from several
		processes.  In most cases, the value measured at a scale $\mu$
		has been evolved to $\mu = M_{Z}$.  Error bars include the
		theoretical uncertainties. 
        Adapted from Ref.~\onlinecite{Bethke:2009jm}.}
	\label{fig:asMZ}
\end{figure}

When evolved to a common scale $\mu = M_{Z}$, the various 
determinations of $\alphas$ lead to consistent values, as shown in 
Fig.~\ref{fig:asMZ}.
A representative mean value (Ref.~\onlinecite{Bethke:2009jm}) is 
\begin{equation}
    \alphas(M_{Z}) = 0.1184 \pm 0.0007 .
    \label{eq:sbalph}
\end{equation}

The agreement of the determination of $\alphas$ from the hadron 
spectrum, via lattice QCD, and from high-energy scattering processes, 
via perturbative QCD (and factorization for deeply inelastic scattering), 
indicates that QCD describes both hadron and partons.
In other words, a single theory accounts for all facets of the strong 
interactions.

\subsection{Hadron spectrum}
\label{sec:hadronmasses}

Soon after the conception of quantum chromodynamics, theorists 
formulated QCD-inspired models to open a dialogue with experiment.
Simple notions about the order of levels, augmented by an effective 
color-hyperfine interaction were put forward in
\citentry{DeRujula:1975ge}
\noindent The extension to excited baryons was given by
\citentry{Isgur:1978xj}
\noindent An extensive analysis of the meson spectrum in a QCD-inspired 
quark model is
\citentry{Godfrey:1985xj}

Massless quarks were confined within a finite radius 
by fiat in the MIT bag model, which is explained in
\citentry{Johnson:1979ni}
\citentry{DeTar:1983rw}

\begin{figure}[bt] 
	\includegraphics[width=8.5cm]{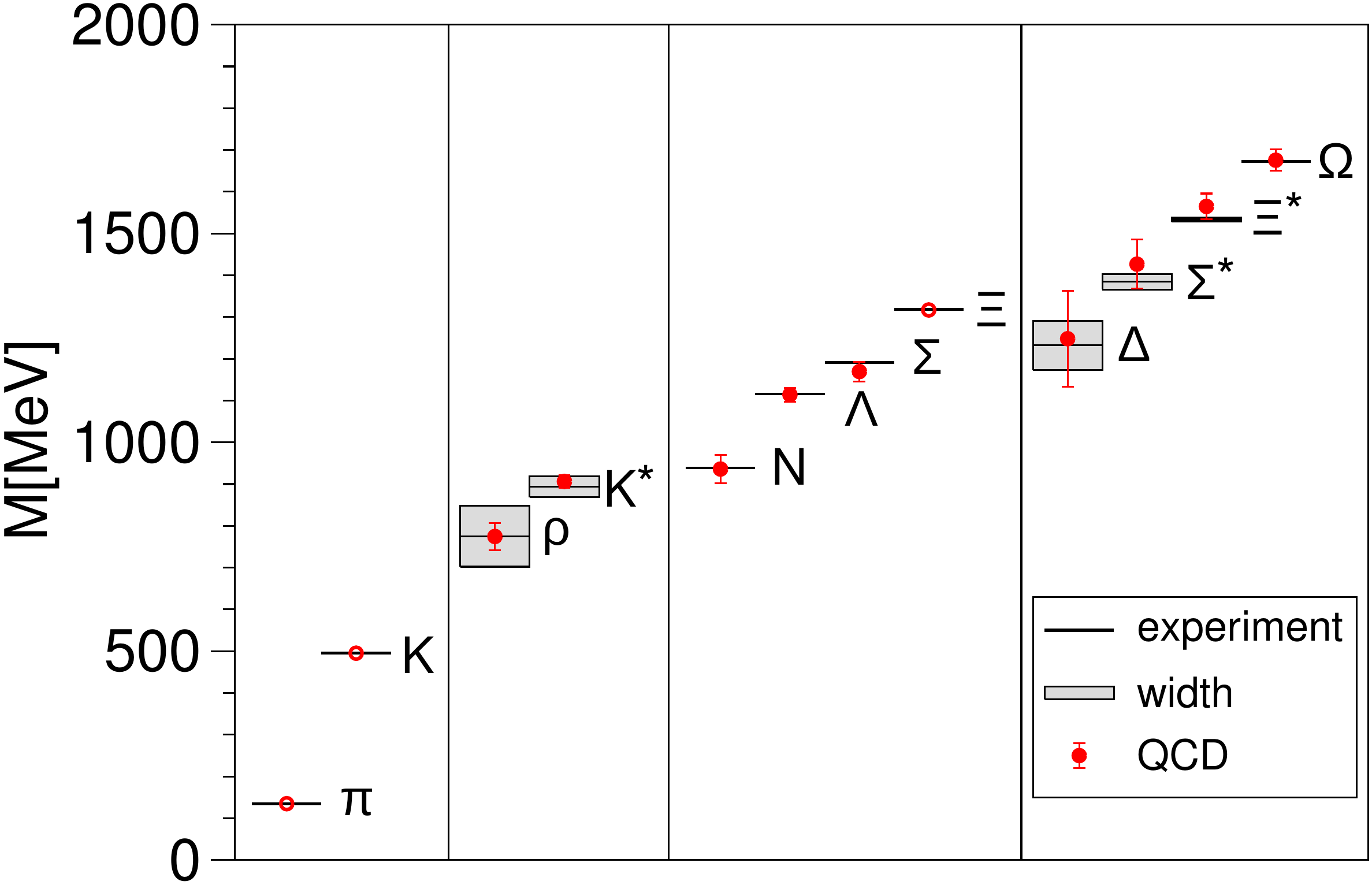}
	\vspace{10pt}
	\caption{Meson ($\pi$, $K$, $\rho$, $K^*$) and baryon masses 
    (others) from lattice QCD, using the $\pi$, $K$, and $\Xi$ masses
    to fix the three free parameters ($m_u+m_d$, $m_s$, 
    $\Lambda_{\rm QCD}$). 
    From Ref.~\onlinecite{Durr:2008zz}.}
	\label{fig:hadronmasses}
\end{figure}

Lattice QCD provides a way to compute the hadron mass spectrum directly 
from the QCD Lagrangian.
The state of the art for light hadrons is shown in 
Fig.~\ref{fig:hadronmasses} and described in
\citentry{Durr:2008zz}
\citentry{Aoki:2008sm}
\citentry{Bazavov:2009bb}

With $\alphas$, the quark masses are the fundamental 
parameters of QCD.
Hadron masses depend on the quark masses, so these calculations yield 
as by-products the best estimates of the light-quark masses
(Ref.~\onlinecite{Bazavov:2009bb})
\begin{eqnarray}
    m_u & = & 1.9 \pm 0.2\mev, \nonumber \\
    m_d & = & 4.6 \pm 0.3\mev, \label{eq:MILC} \\
    m_s & = & 88 \pm 5\mev; \nonumber
\end{eqnarray}
or, defining $\widehat{m}=(m_u + m_d)/2$,
\begin{eqnarray}
    \widehat{m} & = & 3.54^{+0.64}_{-0.35}\mev,
    \nonumber \\
    m_s & = & 91.1^{+14.6}_{-6.2}\mev,
    \label{eq:JLQCD}
\end{eqnarray}
from
\citentry{Ishikawa:2007nn}
\noindent Both groups use $2+1$ flavors of sea quarks.
The quoted masses are in the $\overline{\mathrm{MS}}$ scheme at $2\gev$.
Ratios of these results agree with chiral perturbation theory:
\citentry{Weinberg:1977hb}
\citentry{Gasser:1982ap}
\citentry{Donoghue:1989sj}

The estimates Eqs.~(\ref{eq:MILC}) and~(\ref{eq:JLQCD}) show that the 
up- and  down-quark masses account for only 
$3\widehat{m} \approx 10\mev$ out of the nucleon mass of $940\mev$.
Accordingly, to percent-level accuracy, nearly all the mass of everyday 
matter arises from chromodynamic energy of gluons and the kinetic 
energy of the confined quarks.

In the elementary quark model, mesons are $q\bar{q}$ color 
singlets, whereas baryons are $qqq$ color singlets.
Although QCD favors these configurations as the states of lowest 
energy, it also admits other body plans: quarkless mesons called 
glueballs, $q\bar{q}g$ mesons called hybrids, $qq\bar{q}\bar{q}$ mesons 
called tetraquarks, $qqqq\bar{q}$ baryons called pentaquarks, etc.
At this time, there are no credible reports of non--quark-model baryons. 
The rich body of experimental information on  non--quark-model mesons 
is reviewed in
\citentry{Klempt:2007cp}
\citentry{Crede:2008vw}
\noindent Strong theoretical evidence for glueballs comes from 
lattice-QCD calculations in an approximation to QCD without quarks:
\citentry{Morningstar:1999rf}
\citentry{Sexton:1995kd}
\citentry{Meyer:2004jc}

\subsection{\boldmath The reaction $e^+e^-\to\textrm{hadrons}$}
\label{sec:Rmumu}

In the framework of the quark-parton model, the cross section for
hadron production in electron-positron annihilations at
center-of-momentum energy $\sqrt{s}$ is given by 
\begin{equation}
	\sigma_{\mathrm{qpm}}(e^+e^- \to \mathrm{hadrons}) =
		\frac{4\pi\alpha^2}{3s}\left[
		3 \sum_q e_q^2 \theta(s - 4m_q^2) \right],
	\label{eq:Rmumu}
\end{equation}
where $e_q$ and $m_q$ are the charge and mass of quark flavor $q$ and
the step function $\theta$ is a crude representation of kinematic 
thresholds.

\begin{figure}[bt] 
	\includegraphics[width=8.5cm]{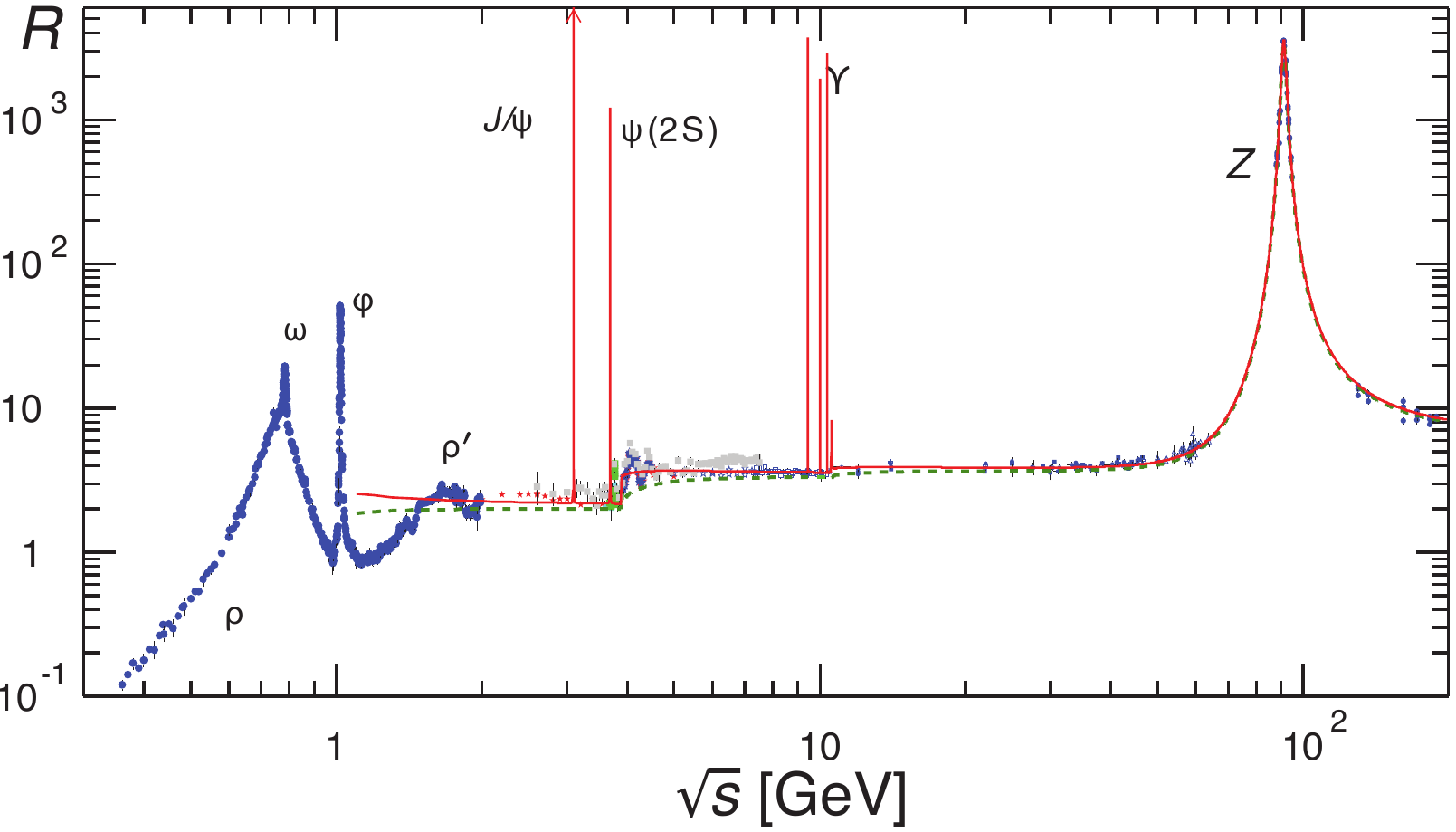}
	\vspace{10pt}
	\caption{World data on the ratio $R$ from \Eqn{eq:Rratio}, compared 
        with predictions of the quark-parton model (dashed curve) and
		perturbative QCD at three loops (solid line), from
		Ref.~\onlinecite{Amsler:2008zzb}.}
	\label{fig:Rmumu}
\end{figure}

The factor 3 preceding the sum over active flavors is a
consequence of quark color. The rough agreement between measurements of
the ratio of hadron production to muon-pair production and the
prediction (\ref{eq:Rmumu}), shown as the dashed line in
Fig.~\ref{fig:Rmumu}, is powerful evidence that quarks are color
triplets.

The parton-level prediction is modified by real and virtual emission of
gluons, much as the quantum electrodynamics prediction for
$\sigma(e^+e^- \to \mu^+\mu^-) = {4\pi\alpha^2}/{3s}$ is changed by
real and virtual emission of photons. To leading order in the running
coupling $\alphas(s)$, the result is
\begin{equation}
    \sigma_{\mathrm{QCD}}(e^+e^- \to \mathrm{hadrons}) =
    \sigma_{\mathrm{qpm}}\left[ 1 + \frac{\alphas}{\pi} +
    \mathcal{O}(\alphas^2) \right].
    \label{eq:RinQCD}
\end{equation}
The QCD prediction for
\begin{equation}
    R \equiv \frac{\sigma(e^+e^-\to\mathrm{hadrons})}%
        {\sigma(e^+e^- \to \mu^+\mu^-)},
    \label{eq:Rratio}
\end{equation}
now known through order $\alphas^3$, is shown as the solid 
line in Fig.~\ref{fig:Rmumu}. 

The success of the perturbative prediction hangs on the validity of
asymptotic freedom, to be sure, but also on the utility of
quark-hadron duality and inclusive nature of the total hadronic cross
section, such that potential infrared divergences cancel.
The calculational technology, for the case of negligible quark masses, 
is reviewed in
\citentry{Chetyrkin:1996ia} 
\noindent Many studies of QCD in the reaction $e^+e^- \to Z$ are 
reviewed in 
\citentry{Bethke:1992gh}

Moments of the cross section
\begin{equation}
    \mathcal{M}_n\equiv \int_{4m_Q^2}^\infty ds\, s^{-(n+1)}R_Q(s),
\end{equation}
where $R_Q$ is the part of $R$ [\Eqn{eq:Rratio}] due to $Q\bar{Q}$, 
are useful for determining the masses of the charmed and bottom quarks.
The most recent results are
\begin{eqnarray}
    m_c(3\gev) & = & 986\pm13\mev, \\
    m_b(10\gev) & = & 3610\pm16\mev,
\end{eqnarray}
where the values are again in the $\overline{\mathrm{MS}}$ scheme and the 
argument indicates the renormalization point.
These results are taken from
\citentry{Chetyrkin:2009fv}
\noindent which also serves as a useful entr\'{e}e to the literature.

\subsection{\boldmath Jets and event shapes in $e^+e^-\to\textrm{hadrons}$}
\label{subsec:shapes}

A \textit{hadron jet} is a well-collimated cone of correlated particles 
produced by the hadronization of an energetic quark or gluon. 
Evidence that hadron jets produced in the electron-positron
annihilation into hadrons  follow the distributions calculated for
$e^{+}e^{-} \rightarrow q\bar{q}$ was presented in
\citentry{Schwitters:1975dm}
\citentry{Hanson:1975fe}

The notion that gluon radiation should give rise to three-jet events
characteristic of the final state $q\bar{q}g$ was made explicit by
\citentry{Ellis:1976uc}
\noindent and confirmed in experiments at the PETRA storage ring at the 
DESY Laboratory in Hamburg:  
\citentry{Brandelik:1979bd}
\citentry{Berger:1979cj}
\citentry{Bartel:1979ut}
\citentry{Barber:1979yr}
\noindent For a retrospective account of the discovery, see
\citentry{Soding:1996zk}

The definition of a three-jet cross section corresponding to the
quark-antiquark-gluon final state is plagued by infrared
difficulties---as is the specification of any final state with a
definite number of partons.
It is, however, possible to define infrared-safe energy-weighted cross 
sections that are calculable within QCD, as shown in
\citentry{Sterman:1977wj}
\noindent Modern definitions of jets---taking infrared safety, 
calculability, ease of measurement into account, and the extension to 
hadronic collisions---are surveyed in
\citentry{Salam:2009jx}
\noindent Definitions within SCET are discussed in 
Refs.~\onlinecite{Becher:2009qa,Beneke:2009rj,Chiu:2009mg,%
Stewart:2009yx,Mantry:2009qz,Ellis:2009wj}.

\begin{figure}[tb] 
	\includegraphics[width=8.5cm]{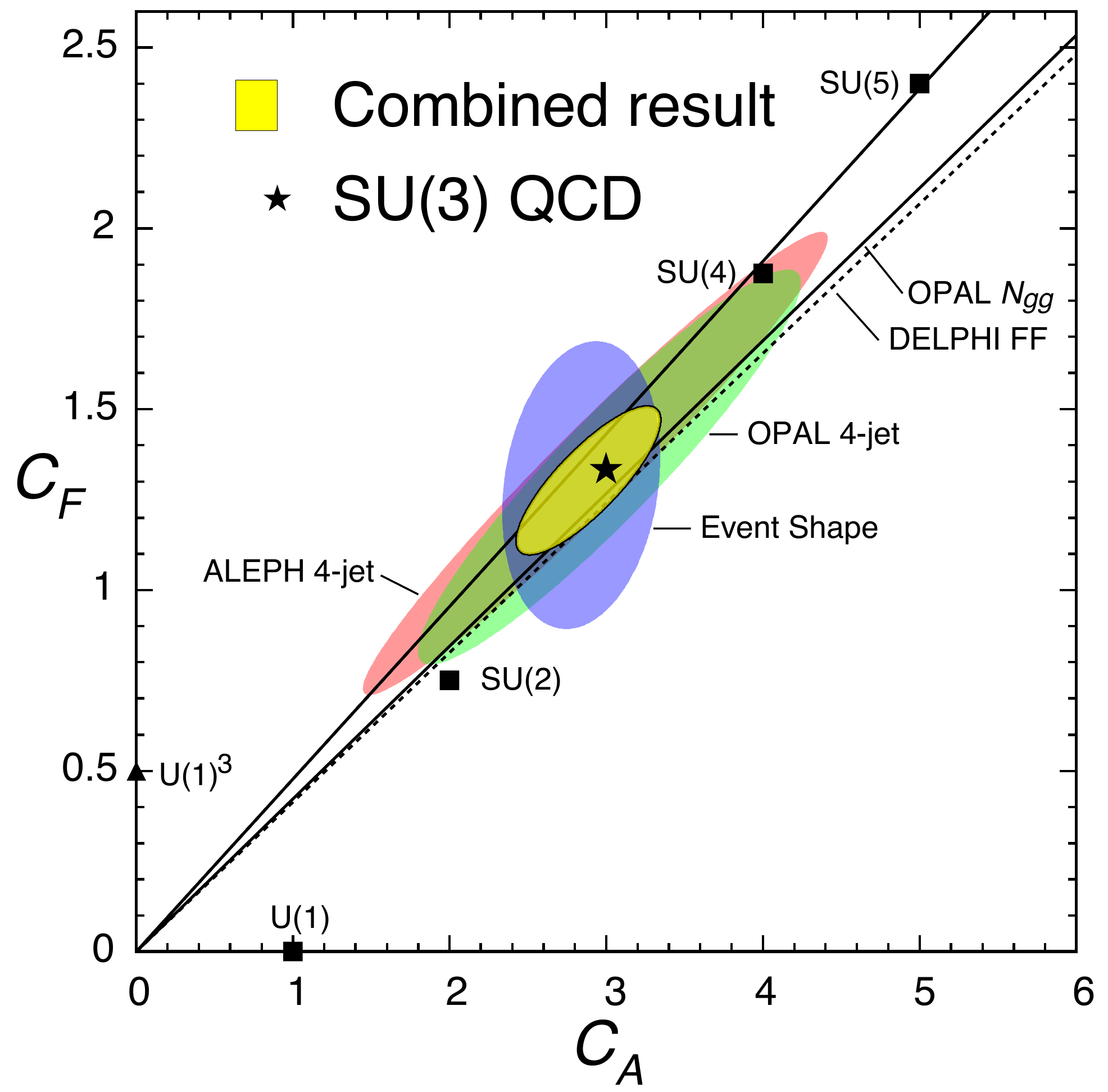}
	\vspace{10pt}
	\caption{Determinations of the color factors $C_F$ and $C_A$ from 
    studies of event shapes, jet cross sections, and fragmentation 
    functions at LEP (adapted from Ref.~\onlinecite{Kluth:2006bw}). 
    The common fit, shown as the shaded ellipse, favors the QCD 
    expectations indicated by {\LARGE$\star$}, and discriminates 
    against other hypotheses.
    The error ellipses show the 90\% confidence level.}
	\label{fig:CFCA}
\end{figure}

Various observables are sensitive to different combinations of 
the quark and gluon color factors, $C_F$ and $C_A$, and so an ensemble 
of measurements may serve to test the QCD group-theory structure via 
\Eqn{eq:qgcolorf}.
The constraints from a number of studies at LEP are 
compiled in Fig.~\ref{fig:CFCA}.
The combined result, presented in
\citentry{Kluth:2006bw}
\noindent yields
\begin{eqnarray}
    C_F & = & 1.30 \pm 0.01\hbox{ (stat.)} \pm 0.09\hbox{ (syst.)}\nonumber \\
    C_A & = & 2.89 \pm 0.03\hbox{ (stat.)} \pm 0.21\hbox{ (syst.)} ,
    \label{eq:expqgcolorf}
\end{eqnarray}
in excellent agreement with the expectations $C_F = \cfrac{4}{3}$, $C_A = 3$.

\subsection{Departures from Bjorken scaling in deeply inelastic scattering}
\label{sec:DIS}

At high resolution (high $Q$), the detailed composition of the nucleon is 
described by the parton distribution functions extracted in deeply 
inelastic lepton-nucleon scattering and other hard processes.
The consequences for lepton-nucleon scattering were first worked out
(to leading order) in 
\citentry{Gross:1973ju}
\citentry{Georgi:1951sr}

According to the parton model, a hadron is a collection of quasifree
quarks, antiquarks, and gluons.
In the ``infinite momentum frame,'' in which the longitudinal momentum 
of the hadron is very large, each parton carries a fraction $x$ of the 
hadron's momentum.
A~parton distribution function $f_i(x_i)$ specifies the probability of 
finding a parton of species $i$ with momentum fraction $x_i$. 
A highly intuitive formalism that generalizes the parton distributions
to $f_i(x_i,Q^2)$ and stipulates the evolution of parton distributions
with momentum transfer $Q^2$ was given in
\citentry{Altarelli:1977zs}
\citentry{Gribov:1972ri}
\citentry{Dokshitzer:1977sg}
\noindent The Altarelli-Parisi prescription is appropriate for moderate 
values of $x$ and large values of $Q^2$.
The extension to higher-order corrections in the 
$\overline{\mathrm{MS}}$ scheme is presented in
Ref.~\onlinecite{Bardeen:1978yd} and reviewed in
\citentry{Buras:1979yt}
\noindent An early quantitative test appears in 
\citentry{Abramowicz:1982jd}

Increasingly comprehensive data sets deepened the dialogue between
theory and experiment. For an informative sequence of reviews, see
\citentry{Drees:1983pd}
\citentry{Mishra:1989jc}
\citentry{Roberts:1990ww}
\citentry{Conrad:1997ne}  
\citentry{Abramowicz:1998ii}
\citentry{Devenish:2004pb}
\citentry{Klein:2008di}
\noindent The series of annual workshops on deeply inelastic scattering
and QCD may be traced from
\citentry{DIS09}

In addition to its ``valence'' components, a hadron contains 
quark-antiquark pairs and gluons, by virtue of quantum fluctuations.
In the extreme limit $Q \to \infty$, for any hadron, the momentum 
fraction carried by gluons approaches 8/17, and that carried by any of 
the six species of quark or antiquark approaches~3/68.
The asymptotic equilibrium partition reflects the relative strengths of 
the quark-antiquark-gluon and three-gluon couplings, as well as the 
number of flavors.
The current state of the art for parton distributions (at finite $Q$) is 
comprehensively documented in
\citentry{Dittmar:2009ii}
\noindent A library providing a common interface to many modern sets of 
parton distributions is
\citentry{LHAPDF}
\noindent The sets of parton distributions currently in wide use may be
traced from
\citentry{Nadolsky:2008zw}
\citentry{CTEQweb}
\citentry{Martin:2009iq}
\citentry{MSTWweb}
\citentry{JimenezDelgado:2008hf}
\citentry{GRVweb}
\citentry{H1:2009wt}
\citentry{Ball:2010de}
\citentry{Alekhin:2006zm}

It is conventional to separate quark (and antiquark) distributions into
``valence'' components that account for a hadron's net quantum numbers
and ``sea'' contributions in which quarks balance antiquarks overall.
Neither a symmetry nor QCD dynamics demand that $q_i(x) = \bar{q}_i(x)$
locally, and experiment has now revealed a flavor asymmetry in the
light-quark sea of the proton.
\citentry{McGaughey:1999mq} 
\citentry{Kumano:1997cy}
\noindent Sum rules that parton distributions must respect in QCD are
reviewed in
\citentry{Hinchliffe:1996hc}

The number densities $q(x,Q^2)$, $\bar{q}(x,Q^2)$, and $g(x,Q^2)$, of 
quarks, antiquarks, and gluons within a hadron can be calculated at 
large $Q^2$ by Altarelli-Parisi evolution 
(Ref.~\onlinecite{Altarelli:1977zs}) from initial distributions 
determined at $Q_0^2$.
However, at small values of the momentum fraction $x$, the resulting
densities may become large enough that the partons overlap spatially,
so that scattering and recombination may occur, as argued in
\citentry{Gribov:1984tu}
\noindent Recombination probabilities were computed in
\citentry{Mueller:1985wy}
\noindent and expectations for lepton-nucleon scattering at very small 
values $x$ are developed in
\citentry{Badelek:1992gs}
\citentry{Badelek:1994fe}
\citentry{Lipatov:1997ts}
\citentry{Levin:1999mw}
\noindent Experiments at the $e^\mp p$ collider HERA, which operated at 
c.m.\ energies up to $\sqrt{s} = 320\gev$, probed the small-$x$ regime 
and established a rapid rise in the parton densities as $x \to 0$, as 
reviewed in Refs.~\onlinecite{Abramowicz:1998ii,Klein:2008di}.
However, recombination phenomena have not yet been demonstrated.
Implications of the HERA observations for future experiments are 
explored in 
\citentry{Frankfurt:2005mc}

Our knowledge of the spin structure of the proton at the constituent
level is drawn from polarized deeply inelastic scattering experiments, 
in which polarized leptons or photons probe the structure of a 
polarized proton and polarized proton-proton collisions.
How current understanding developed, and what puzzles arose, can 
be traced in
\citentry{Lampe:1998eu}
\citentry{Hughes:1999wr} 
\citentry{Filippone:2001ux}
\citentry{Bass:2004xa}
\citentry{Vogelsang:2007zza}
\citentry{Bass:2007zzb}
\citentry{Thomas:2008bd}
\citentry{Kuhn:2008sy}
\citentry{Bradamante:2008zza}
\noindent Progress in making spin-dependent measurements can be traced 
through the spin physics symposia; the latest in the series is
\citentry{Spin08}

For a set of spin-dependent parton distribution functions, with 
extensive references to the underlying measurements, see
\citentry{deFlorian:2009vb}

Standard parton distribution functions provide detailed information
about how spin and longitudinal momentum and spin are partitioned among
the quarks, antiquarks, and gluons in a fast-moving hadron, but the
information is integrated over transverse degrees of freedom.
The role of orbital angular momentum of the partons in building a
spin-$\frac{1}{2}$ proton is obscured.
Generalized parton distributions inferred from exclusive scattering 
processes provide a tool for probing such subtleties of hadron structure.
\citentry{Diehl:2003ny}
\citentry{Ji:2004gf}
\citentry{Belitsky:2005qn}
\citentry{Hagler:2009ni}

An important undertaking of modern hadron physics is to understand 
how hidden flavors (\eg, virtual $s\bar{s}$ pairs) contribute to the 
structure of the nucleon.
Recent experimental and theoretical progress toward unravelling the 
role of strange quarks in the nucleon can be traced in
\citentry{Arrington:2006zm}
\citentry{Perdrisat:2006hj}

In analogy to the hidden flavors of light quarks, hadrons could have an 
intrinsic component of charm-anticharm pairs:
\citentry{Brodsky:1980pb}
\citentry{Pumplin:2007wg}
\citentry{Alekhin:2009ni}

\subsection{Quarkonium}

An early opportunity for QCD-inspired models of hadrons came with the
discovery of the $J/\psi$ particle and other bound states of
charmed quarks and antiquarks,
\citentry{Aubert:1974js}
\citentry{Augustin:1974xw}
\noindent For an account hard on the heels of the discovery, see
\citentry{Drell:1975cf}
\noindent Early perspectives on the implications are given in the Nobel 
Lectures,
\citentry{Ting:1977xk}
\citentry{Richter:1977fn}

Quarkonium spectroscopy was enriched by the discovery of the $\Upsilon$
family of $b\bar{b}$ bound states:
\citentry{Herb:1977ek}
\citentry{Innes:1977ae}
\noindent An accessible account of these discoveries is
\citentry{Lederman:1978gi}
F\noindent or a summary of early comparisons between the $c\bar{c}$ and
$b\bar{b}$ families, see
\citentry{Bloom:1982rc}

These discoveries spurred the development of potential models 
(see Sec~\ref{sec:QQ}).
Reviews of this work from the experimental perspective are in
\citentry{Franzini:1984nq}
\citentry{Besson:1993mm}

Calculations of the quarkonium spectrum, once the exclusive province of 
potential models (\cf\ Sec.~\ref{sec:QQ}), are an important theme in 
lattice QCD.
Three of the first papers on calculations with $2+1$ flavors of sea 
quarks are
\citentry{Gray:2005ur}
\citentry{Follana:2006rc}
\citentry{Burch:2009az}

The breadth of quarkonium physics---experimental, theoretical, and 
computational---is surveyed in
\citentry{Brambilla:2004wf}

A novel form of quarkonium arises from binding a bottom quark and a 
charmed antiquark.
The first observation of the pseudoscalar $B_c$ meson is reported in
\citentry{Abe:1998wi}
\noindent Precise measurements of the mass did not appear until later:
\citentry{Abulencia:2005usa}
\noindent The mass of the $B_c$ was correctly predicted by PNRQCD:
\citentry{Brambilla:2000db}
\noindent and lattice QCD:
\citentry{Allison:2004be}

Recently, a new set of states has appeared in the charmonium spectrum 
that presents new challenges to hadron dynamics.
Some of these may be (mostly) charm-anticharm states above the 
threshold for decay into charmed-meson pairs. 
Others cannot readily be identified in the same way.
For a recent survey, see  
\citentry{Godfrey:2008nc} 

\subsection{Jets in hadron collisions}

An account of early evidence for jet structure in the first $pp$ 
collider, at energies up to $\sqrt{s}=63\gev$, is given in
\citentry{Akesson:1983rk}
\noindent Incisive comparisons with QCD were made in experiments at the 
SPS Collider, at energies up to $630\gev$:
\citentry{Appel:1985rm}
\citentry{Alitti:1990aa}
\citentry{Arnison:1986vk}

\begin{figure}[tb] 
	\includegraphics[width=8.5cm]{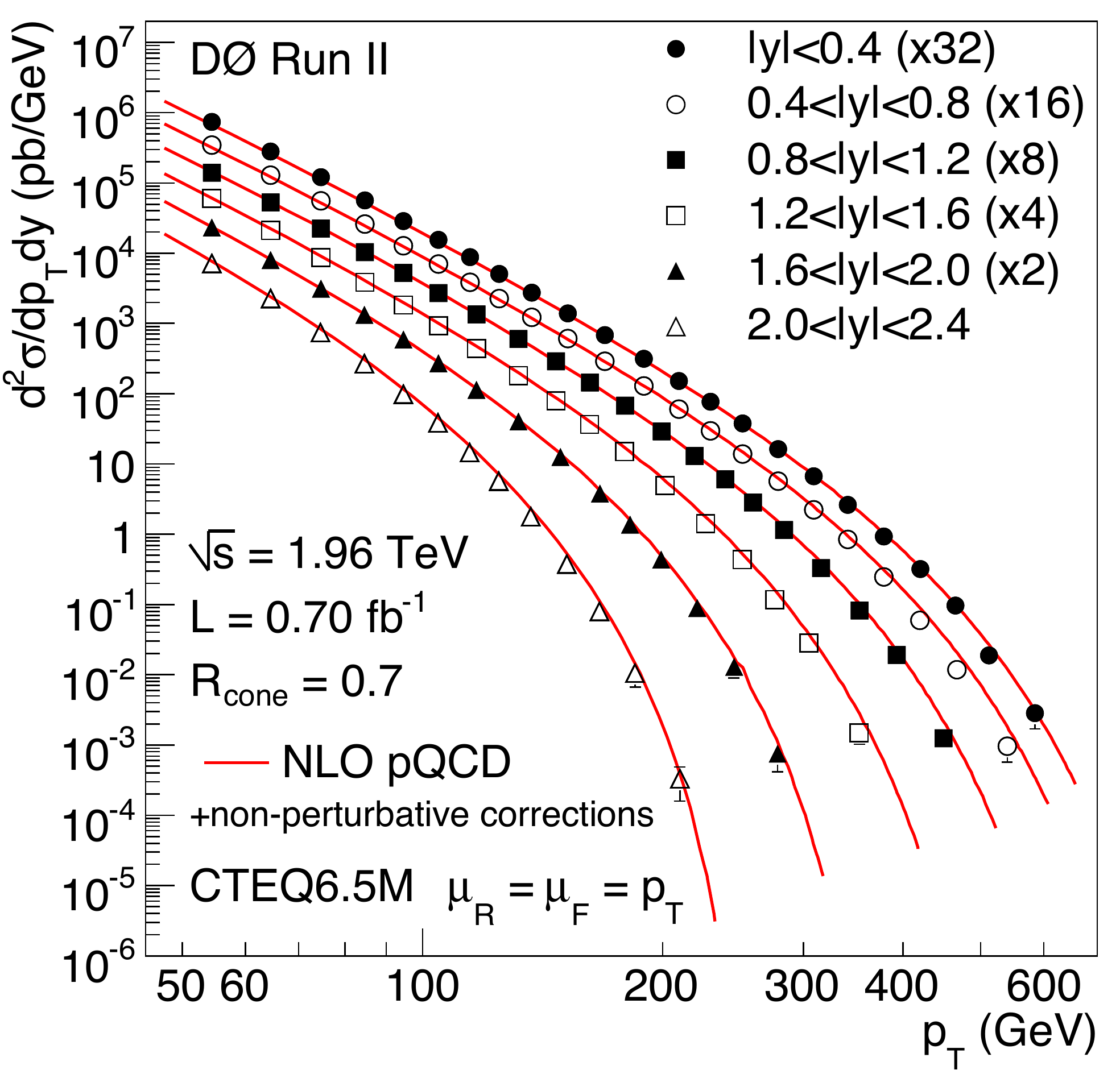}
	\vspace{10pt}
	\caption{The inclusive jet cross section measured at $\sqrt{s} = 
        1.96\tev$ by the D0 Collaboration (Ref.~\onlinecite{:2008hua}) 
        as a function of transverse momentum in six rapidity bins. 
        The data points are multiplied by 2, 4, 8, 16, and 32
        for the bins $1.6<|y|<2.0$, $1.2<|y|<1.6$, $0.8<|y|<1.2$, 
        $0.4<|y|<0.8$, and $|y|<0.4$, respectively.
        Solid curves show next-to-leading-order (one-loop)
		perturbative QCD predictions.
		A $6.1\%$ uncertainty on the integrated luminosity is not
		included.
        Theoretical predictions carry an uncertainty of 
        approximately~$10\%$.}
	\label{fig:d0jets}
\end{figure}

Extensive studies have been carried out at the Tevatron Collider, at 
energies up to $\sqrt{s} = 1.96\tev$.
We show in Fig.~\ref{fig:d0jets} that perturbative QCD, 
evaluated at next-to-leading order, accounts for the 
transverse-momentum spectrum of central jets produced in the reaction
\begin{equation}
    \bar{p}p \to \hbox{jet}_{1} + \hbox{jet}_{2} + \hbox{anything}
    \label{eq:pbarpjj}
\end{equation}
over more than eight orders of magnitude:
\citentry{:2008hua}
\noindent Similar results from the CDF experiment are reported in
\citentry{Abulencia:2007ez}
\noindent For a summary of recent QCD studies at the Tevatron, see
\citentry{Bhatti:2010bf}
\noindent The current state of the art is presented in
\citentry{Campbell:2006wx}
\noindent Jet phenomena in relativistic heavy-ion collisions are 
summarized in
\citentry{Tannenbaum:2006ku}
\noindent For a discussion of jet definitions and their interplay with 
measurements, see Ref.~\onlinecite{Salam:2009jx}.

\subsection{Photon structure function}

The proposal to determine the constituent structure of the photon by
studying the scattering of a highly virtual photon on a real photon is
due to
\citentry{Brodsky:1971vm}
\citentry{Walsh:1971xy}
\noindent To the extent that a photon behaves as a vector meson, the
momentum-fraction ($x$) and momentum-transfer ($Q^2$) dependences of
its structure function should roughly resemble those of the proton
structure function. But a parton-model calculation reveals that a
pointlike contribution that arises when the photon fluctuates into a
quark-antiquark pair should dominate over the vector-meson component at
high $Q^2$.
\citentry{Walsh:1973mz}
\citentry{Kingsley:1973wk}
\noindent Remarkably, the $x$-dependence of the photon structure
function is fully calculable at large $Q^2$, in contrast to the proton
structure function, for which the $x$-dependence at fixed $Q^2$ results
from nonperturbative effects and, in practice, is taken from the data 
or, in the approach of Ref.~\onlinecite{JimenezDelgado:2008hf}, 
from a simple Ansatz.

QCD confirms the calculability of the photon structure function at 
large $Q^2$, and differs from the parton-model result, particularly
as $x\to1$.
In leading logarithmic approximation, the result is reported in
\citentry{Witten:1977ju}
\noindent The next-to-leading-order calculation improves the
reliability of the predicted shape of the photon structure function:
\citentry{Bardeen:1978hg}
\noindent also enabling a determination of the strong coupling $\alphas$,
now at the 5\% level.

For an excellent short review, see 
\citentry{Buras:2005nj}
\noindent Extensive experimental summaries appear in
\citentry{Nisius:1999cv}
\citentry{Krawczyk:2000mf}
\noindent A useful digest appears in Fig.~16.14 of 
Ref.~\onlinecite{Amsler:2008zzb}.

\subsection{Diffractive Scattering}

The Pomeranchuk singularity, or Pomeron, designates the Regge pole with 
vacuum quantum numbers that controls the asymptotic behavior of elastic 
and total cross sections.
The Regge intercept of the Pomeron, the location of the pole in the 
complex angular-momentum plane at zero momentum transfer, would be 
$\alpha_{\mathbb{P}} = 1$ if total cross sections approached constants 
at high energies.
Comprehensive modern fits to meson-baryon and especially 
proton-(anti)proton total cross sections initiated in
\citentry{Donnachie:1992ny}
\noindent indicate that $\alpha_{\mathbb{P}} \approx 1.08$.

With the advent of quantum chromodynamics, it was natural to begin 
searching for a dynamical description of the Pomeron's origin. The idea 
that the Pomeron somehow emerged from the exchange of color-octet 
gluons between color-singlet hadrons, first articulated in
\citentry{Low:1975sv}
\noindent has great resonance today.
A concrete realization of the Pomeron in QCD as a composite state of
two Reggeized gluons was developed (in leading logarithmic approximation)
by
\citentry{Kuraev:1977fs}
\citentry{Balitsky:1978ic}
\noindent The dynamics underlying the ``BFKL Pomeron'' also entail a 
resummation of leading $1/x$ corrections to structure functions.

In spite of much productive effort, the origin of the Pomeron and the 
details of its structure are still not entirely clear:
\citentry{Forshaw:1997dc}
\citentry{Donnachie:2002en}

Although the Pomeron was conceived to account for ``soft'' scattering, 
the QCD interpretation implies that it should have a partonic 
structure, and thus a ``hard'' component.
For a survey of evidence for a hard component from $e^\mp p$ collisions 
at HERA, see
\citentry{Wolf:2009jm}
\noindent The suggestion that Pomeron exchange should result in large 
rapidity gaps between jets crystallized in 
\citentry{Bjorken:1992er}
\noindent Evidence for events of the suggested character was presented in 
\citentry{Abachi:1994hb}
\citentry{Abe:1994de}

If the Pomeron can be exchanged between color-singlets, then, in 
analogy with two-photon physics or the multiperipheral model, two 
Pomerons can collide and produce collections of particles with net 
vacuum quantum numbers, isolated by large rapidity gaps from other 
particle production in the event, as observed in
\citentry{Aaltonen:2009kg}
\noindent A tantalizing possibility is that the Higgs boson might be 
discovered at the LHC in very quiet events:
\citentry{Martin:2006fx}

\subsection{Weak boson production}

In hadron colliders, an electroweak vector boson can be produced 
directly via fusion of a quark and antiquark.
The cross section depends on the parton distributions discussed in 
Sec.~\ref{sec:DIS}.
This extension of the parton model was first noted in
\citentry{Drell:1970wh}
\noindent This process provided the basis for the discovery of the $W$ 
and $Z$ bosons:
\citentry{Rubbia:1985pv}

What was then a QCD-guided discovery is now one of the most precise 
tests of perturbative QCD.
The production cross sections and rapidity distributions for the 
Tevatron and the LHC have been carried out to the 
next-to-next-to-leading order in~$\alphas$ in
\citentry{Hamberg:1990np}
\citentry{Anastasiou:2003ds}
\noindent These calculations have been validated (except at the largest 
accessible values of rapidity) in measurements performed at the Tevatron:
\citentry{Abazov:2007jy}
\citentry{Aaltonen:2009pc}
\noindent Production of electroweak bosons together with jets is 
discussed in Sec.~\ref{sec:top}.

This history is set to repeat itself in the search for the Higgs boson, 
which relies on the next-to-next-to-leading order QCD calculation:
\citentry{Harlander:2002wh}
\citentry{Anastasiou:2004xq}
\citentry{Catani:2007vq}
\noindent The Higgs-boson searches at the Tevatron and the LHC rely on 
these results, and on comparably precise calculations of background 
processes, in an essential way.

\subsection{Heavy-quark production}
\label{sec:top}

Another probe of the short-distance dynamics of QCD is the production 
of heavy quark-antiquark pairs in hadron collisions:
\citentry{Nason:1987xz}
\citentry{Beenakker:1988bq}
\citentry{Mangano:1991jk}
\noindent An important application is the production of the top quark at the 
Tevatron:
\citentry{Laenen:1993xr}
Measurements of the top quark mass have been combined into an average:
\citentry{:2009ec}
\noindent yielding the result
\begin{equation}
    m_t=173.1\pm1.3\gev,
\end{equation}
where this mass has a more conventional definition (similar to that 
of the electron).
The top-quark mass is now precise enough that the ambiguities raised in 
Refs.~\onlinecite{Beneke:1994sw,Bigi:1994em} are becoming 
quantitatively important.

QCD calculations are important not only to gain an understanding of the 
experimental signal, but also to understand the background, which stems 
from $W$ production:
\citentry{Berends:1990ax}
\citentry{Campbell:2002tg}
\citentry{Ellis:2009bu}
\citentry{Berger:2009ep}

\subsection{Inclusive $B$ decays}

Another useful application of perturbative QCD is to inclusive decays 
of hadrons containing a heavy quark.
In practice, this approach applies to hadrons with the bottom quark.
One again appeals to quark-hadron duality and applies the 
operator-product expansion to factorize the differential rate into
short- and long-distance contributions.
This rich subject launched with
\citentry{Chay:1990da}
\noindent The arc of this research is explained pedagogically in 
Ref.~\onlinecite{Manohar:2000dt} and in further detail in
\citentry{Benson:2003kp}

This formalism has several applications, using the experimental data to 
gain insight into long-distance QCD on the one hand, and to determine 
the bottom quark's flavor-changing weak couplings.
Both perspectives are treated in a thorough analysis of the 
then-current theory and data:
\citentry{Buchmuller:2005zv}

A by-product of these analyses is another determination of the 
bottom-quark mass.
The status is summarized in
\citentry{Antonelli:2009ws}
\noindent This review covers all of flavor physics, including aspects 
pertaining to this and the next subsection, and well beyond.

\subsection{Exclusive meson decays}

Pseudoscalar mesons can decay via the weak interaction to a charged 
lepton and its neutrino, and the rate can be compared with lattice-QCD 
calculations of the transition amplitudes.
For $\pi$ and $K$ mesons, the calculations and measurements agree well.
For mesons with heavy quarks, the measurements lag the calculations 
somewhat, and the agreement is good but not spectacular.
The current status is thoroughly discussed in 
Ref.~\onlinecite{Bazavov:2009bb}.

\begin{figure}[bt] 
	\includegraphics[width=8.5cm]{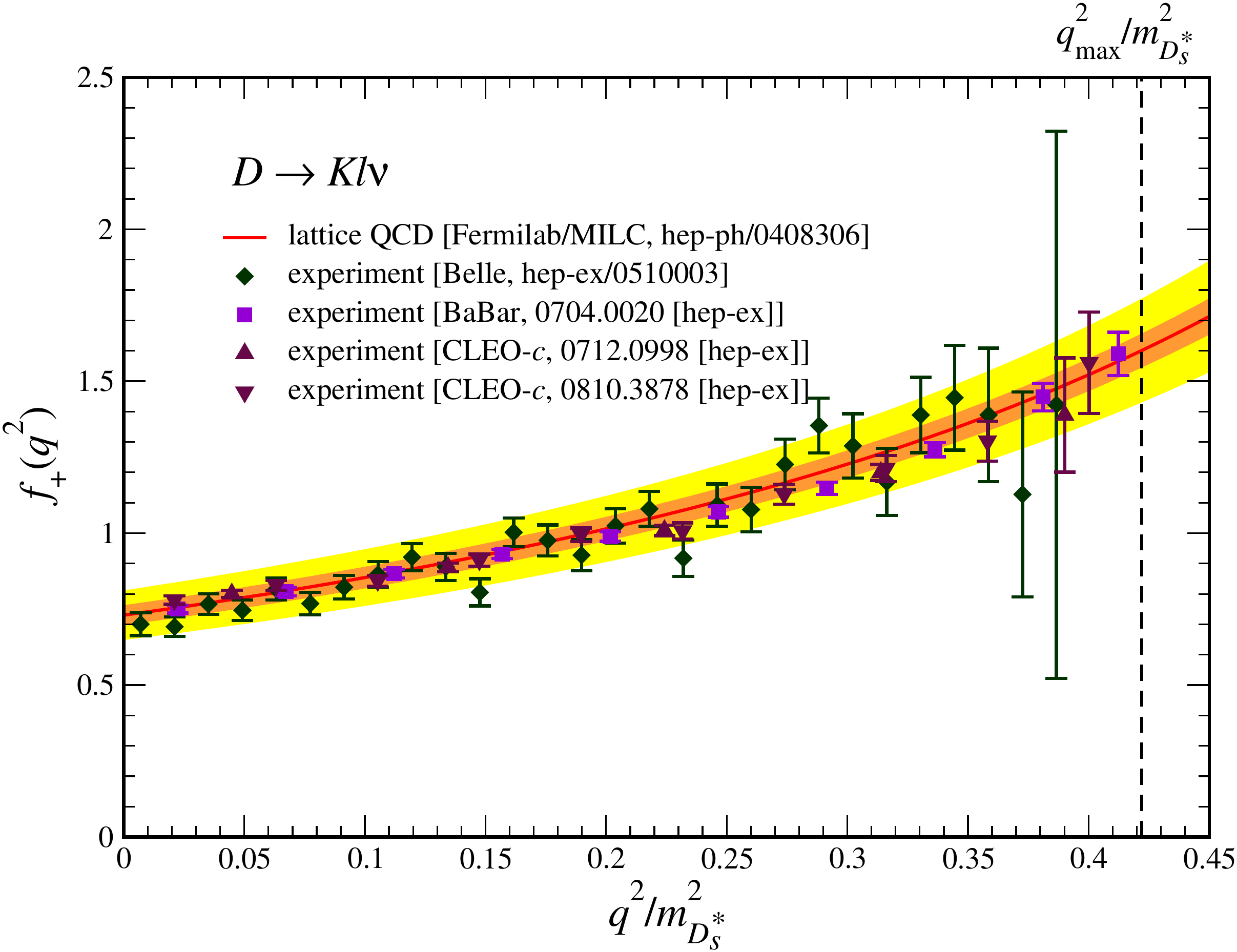}
	\vspace{10pt}
	\caption{Comparison lattice-QCD calculations and experimental 
        measurements of the form factor $f_+(q^2)$ for the semileptonic 
        decay $D\to Kl\nu$.
        The curve and error band show a lattice-QCD calculation;
        From Ref.~\onlinecite{Bazavov:2009bb}; adapted 
        from Ref.~\onlinecite{Bernard:2009ke}.}
	\label{fig:f+DK}
\end{figure}

Pseudoscalar mesons can also decay via the weak interaction to a 
lighter hadron in association with the lepton-neutrino pair.
These three-body decays are called semileptonic.
Lattice-QCD calculations predicted the normalization and kinematic 
distribution of semileptonic $D$ decays.
A good place to start is
\citentry{Bernard:2009ke}
\noindent from which a comparison of a QCD calculation with 
measurements from several experiments is reproduced in 
Fig.~\ref{fig:f+DK}.
Similar comparisons can be made for semileptonic kaon and $B$-meson 
decays.

Nonleptonic kaon decays are too computationally challenging for lattice 
QCD and, apart from constraints from chiral perturbation theory, too 
conceptually challenging via other approaches.
Numerous nonleptonic $B$ decays are kinematically allowed, posing 
conceptual challenges for lattice QCD.
The high scale of the bottom-quark mass, however, allows a 
treatment in perturbative QCD, at least to leading order in $1/m_b$
(Refs.~\onlinecite{Beneke:2000ry},\onlinecite{Bauer:2000yr}).
Broad studies provide information on flavor-changing couplings of the 
Standard Model:
\citentry{Beneke:2001ev}
\citentry{Bauer:2005kd}
\noindent For a comprehensive set of references to the measurements and 
a comparison with calculations at the third order in $\alphas$, 
see Table~3 of
\citentry{Beneke:2009ek}
\noindent For an alternative approach, see
\citentry{Keum:2000wi}

\subsection{Heavy-ion collisions and the quark-gluon plasma}
\label{sec:thermo}

One of the goals of relativistic heavy-ion collisions is to investigate 
the quark-hadron phase transition that presumably occurred in the early 
universe.
\citentry{Riordan:2006df}
\citentry{Brown:1994nk}

By creating small volumes with high energy density or high particle 
density, heavy-ion collisions open a window on new phases of matter.
\citentry{Bjorken:1982qr}
\citentry{Rajagopal:2000wf}
\citentry{Alford:2001dt}
\citentry{Iancu:2003xm}
\citentry{Lappi:2006fp}
\citentry{Alford:2007xm}

Experiments at Brookhaven National Laboratory's 
Relativistic Heavy-Ion Collider (RHIC) imply the 
existence of a ``perfect fluid'' of quarks and gluons.
\citentry{Ludlam:2003rh}
\citentry{Arsene:2004fa}
\citentry{Back:2004je}
\citentry{Adams:2005dq}
\citentry{Adcox:2004mh}
\citentry{Muller:2006ee}
\citentry{BraunMunzinger:2008tz}
\noindent A series of conference on quark matter may be traced starting at
\citentry{QM09}

\begin{figure}[bp]
    \includegraphics[width=8.4cm]{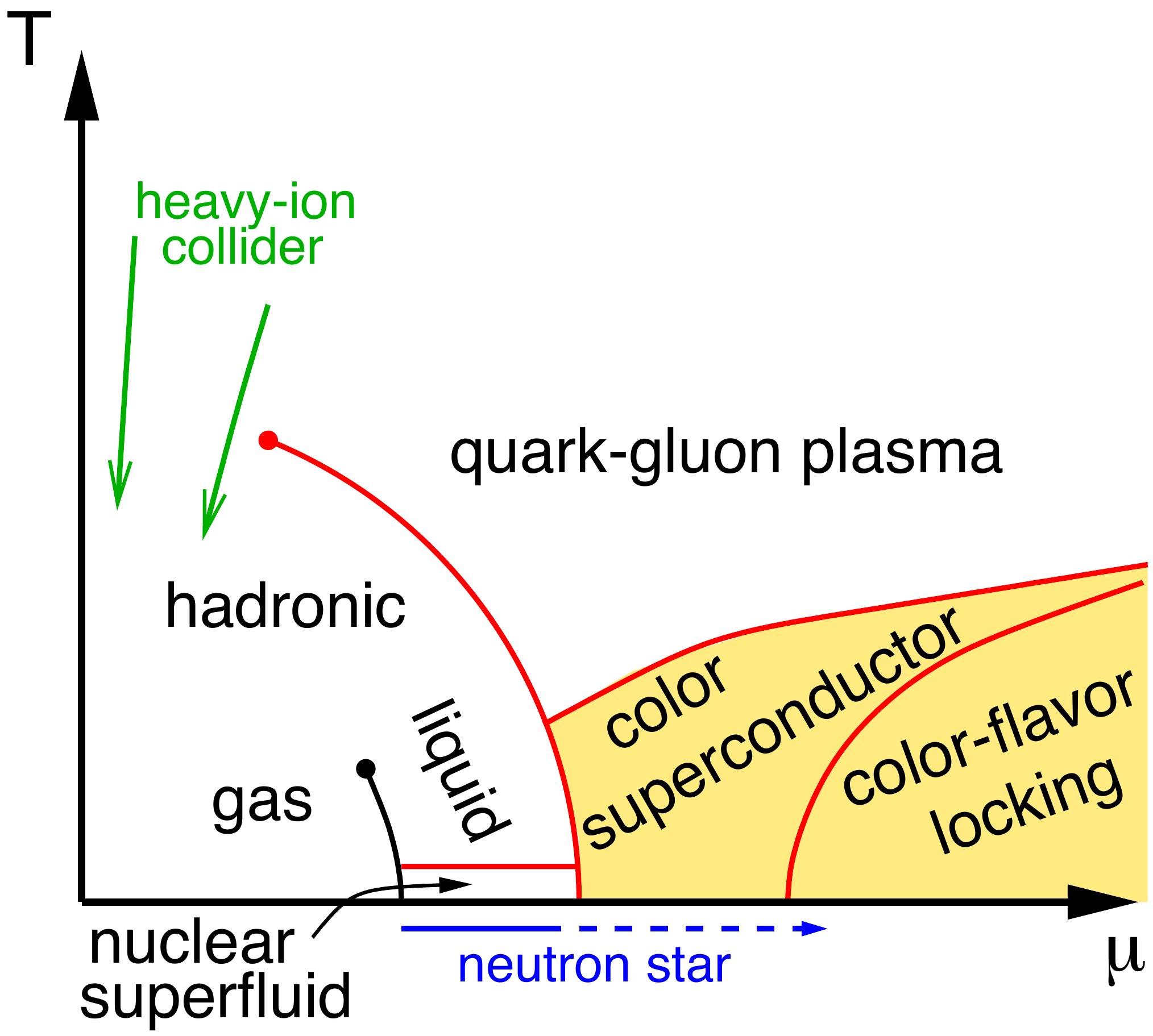}
	\caption{Phase diagram of QCD in the $\mu$-$T$ plane.
    Here $\mu$ denotes baryon chemical potential and $T$ temperature.
    At low $\mu$, there is a smooth transition with varying~$T$, probed 
    by heavy-ion collisions and lattice-QCD calculations.
    At higher~$\mu$ the phases are informed by models and other 
    theoretical considerations.
    Hadronic matter denser than neutron stars is thought to exhibit 
    ``color superconductivity,'' first without and eventually with
    ``color-flavor locking.''
    Adapted from Ref.~\onlinecite{Alford:2007xm}.}
	\label{fig:qcdpd}
\end{figure}

\begin{figure}
    \includegraphics[width=8.5cm]{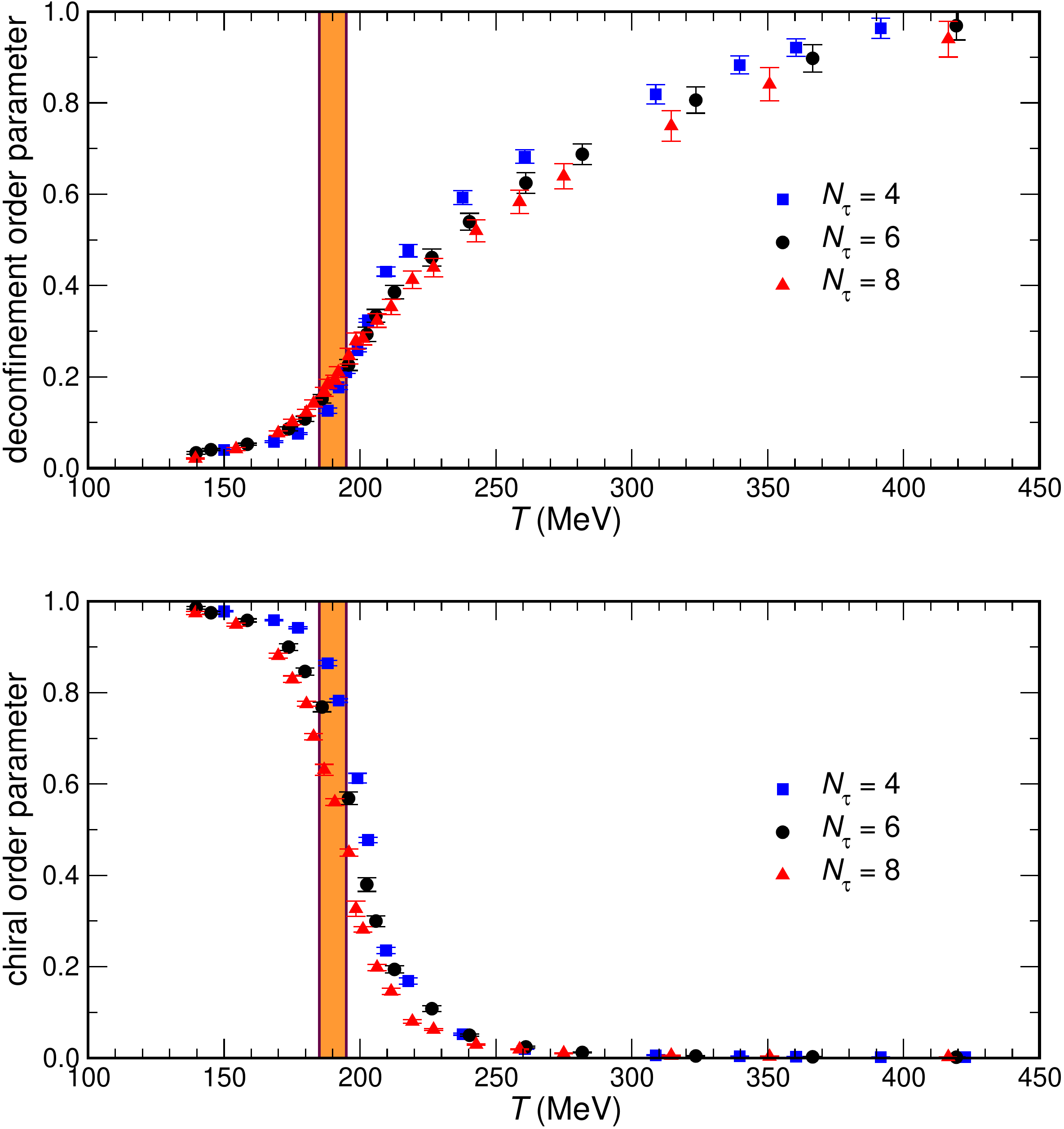}
	\caption{Order parameters for deconfinement (top) and chiral 
        symmetry restoration (bottom), as a function of temperature.
        The physical temperature $T=a/N_\tau$, where $a$ is the lattice 
        spacing.
        Agreement for several values of $N_\tau$ thus indicates that
        discretization effects from the lattice are under control.
        From Refs.~\onlinecite{Cheng:2007jq,Bazavov:2009zn}.}
	\label{fig:Tc}
\end{figure}

The phase diagram is thought to be much richer beyond the region
explored by heavy-ion collisions.
Figure~\ref{fig:qcdpd} shows a current conception of QCD 
thermodynamics (Ref.~\onlinecite{Alford:2007xm}).
The region of Fig.~\ref{fig:qcdpd} with $\mu\approx0$ has been 
demonstrated with lattice QCD:
\citentry{Laermann:2003cv}
\citentry{DeTar:2009ef}
\citentry{Fodor:2009ax}
\noindent In addition to providing detailed information that is useful 
for interpreting heavy-ion collisions, these calculations have shown 
that QCD contains a phase in which (quasiparticle guises of) quarks and 
gluons are no longer confined, and the chiral symmetry of the quarks is 
restored.
As shown in Fig.~\ref{fig:Tc}, the transition is smooth, but order 
parameters for deconfinement and for chiral symmetry restoration change 
qualitatively and quantitatively at essentially the same temperature.
\citentry{Cheng:2007jq}
\citentry{Bazavov:2009zn}
\noindent The transition temperature near 190~MeV corresponds to
$2\times10^{12}$~K.

An intriguing result of these lattice-QCD calculations is how the order 
of the phase transition depends on the light- and strange-quark masses.
If they were around half the size needed to explain the nonzero pion 
and kaon masses, then the transition would be first order, instead of 
a smooth crossover.
\citentry{Aoki:2006we}
\citentry{deForcrand:2006pv}
\noindent That would expose the early universe to a latent heat as it 
cools below the critical temperature.
Ramifications of the QCD phase transition on the early universe are 
discussed in
\citentry{Fuller:1987ue}

Numerical lattice QCD is for now limited to baryon chemical potential 
$\mu\approx0$, with obstacles to the regime relevant to neutron stars.

Ordinary nuclei consist of protons and neutrons, which are composed 
of up and down quarks. Because the most stable nuclei have 
exceedingly long lifetimes---greater than the age of the 
universe---it is natural to idealize them as absolutely stable, up to 
the conjectured nucleon decay that arises in unified theories of the 
strong, weak, and electromagnetic interactions. If the strange-quark 
mass were comparable to the up- and down-quark masses, then the Pauli 
principle would be less restrictive, and the ground state of matter 
would be a mixture of $u$, $d$, and $s$ quarks. It has been 
conjectured that such strange matter is the true ground state in the 
real world, so that nuclear matter is metastable, as elaborated in 
\citentry{Witten:1984rs}
\citentry{Farhi:1984qu}
\noindent Small nuggets of strange matter are called strangelets.

According to the strange-matter hypothesis, compact stars might be 
strange stars, rather than neutron stars, as reviewed in
\citentry{Weber:2004kj}
\noindent which summarizes strange-matter searches.
Conferences on strange quark matter may be traced from
\citentry{SQM09}

Using techniques of gauge-string duality 
(Ref.~\onlinecite{Klebanov:2000me}), theorists have attempted to infer 
characteristics of QCD in the strong-coupling regime from analogue 
theories that possess some degree of supersymmetry.
Applications to heavy-ion collisions and confinement are reviewed in  
\citentry{Son:2007vk}
\citentry{Gubser:2009md}
\noindent One should bear in mind, however, that the archetype of the 
analogue theories, supersymmetric Yang-Mills theory with four 
supercharges, does not share some of the essential features of QCD:
the coupling that corresponds to $\alphas$ does not run, 
and the theory does not confine.

\subsection{QCD and nuclear physics}
\label{sec:nuke}

In principle, all of nuclear physics follows from QCD:
\citentry{Henley:1999fu}
\citentry{Povh:1995mua}
\citentry{Thomas:2001kw}
\noindent A recent review of QCD-based nuclear theory, emphasizing 
symmetries and effective field theories can be found in
\citentry{Epelbaum:2008ga}
\noindent An assault on nuclear physics using lattice gauge theory and 
effective field theory is reviewed in
\citentry{Beane:2008dv}

The nucleon-nucleon potential is governed by pion exchange at distances
beyond $2\fm$, as recognized by
\citentry{Yukawa:1935xg}
\noindent At intermediate range, $1\fm \lesssim r \lesssim 2\fm$, the
nuclear force is determined by the exchange of vector mesons and other
multipion states.
A repulsive hard core, proposed in
\citentry{PhysRev.81.165}
\noindent is essential to the understanding of nuclear stability, the
maximum mass of neutron stars, and other characteristics of nuclear
matter. 
Microscopic quantum Monte Carlo calculations of the properties of light 
nuclei demonstrate that nuclear structure, including both 
single-particle and clustering aspects, can be explained starting from 
elementary two- and three-nucleon interactions
\citentry{Pieper:2001mp}

The essential features of the two-nucleon interaction have now been 
deduced from lattice QCD simulations in an approximation omitting sea 
quarks, as reported in
\citentry{Ishii:2006ec}
\noindent See also the commentary in
\citentry{WilczekHC}
\noindent Meanwhile, this and many other aspects of low-energy 
baryon-baryon interactions have been computed in lattice QCD with $2+1$ 
flavors of sea quarks:
\citentry{Beane:2009py}
 
The European Muon Collaboration (EMC) discovered that the per-nucleon 
deeply inelastic structure function, $F_2(x)$, was significantly 
different for iron than for deuterium, with a marked suppression of 
quarks in the interval $0.3 < x < 0.8$:
\citentry{Aubert:1983xm}
\noindent Data from many subsequent experiments and candidate 
interpretations are reviewed in
\citentry{Geesaman:1995yd}
\citentry{Norton:2003cb}
\noindent Recently, the experimental information has been extended to 
light nuclei:
\citentry{Seely:2009gt}

\section{QCD in the Broader Context of Particle Physics}
\label{sec:beyond}

Quantum chromodynamics is part of the extremely successful 
``Standard Model of Elementary Particles.''
Some resources that help put QCD in the broader context of the Standard 
Model are given here.

A comprehensive source of general knowledge about particle
physics, including many aspects of QCD, is the biannual review by the 
Particle Data Group~ (Ref.~\onlinecite{Amsler:2008zzb}).

Many of the themes that came together in quantum chromodynamics may be 
traced in the contributions to two symposia on the history of particle 
physics:
\citentry{Brown:1989im}
\citentry{Hoddeson:1997hk}

Experimental steps that led to today's standard model of particle
physics are surveyed in the well-chosen collection,
\citentry{Cahn:1989by}
\noindent Also see
\citentry{Glashow:1975pd}
\citentry{'tHooft:1980us}
\citentry{Quigg:1985ai}
\citentry{Close:2007nv}

Like quantum chromodynamics, the electroweak theory is a gauge theory,
based on weak-isospin and weak-hypercharge symmetries described by the
gauge group \ewgg.  For a look back at the evolution of the electroweak
theory, see the Nobel Lectures by some of its principal architects:
\citentry{Weinberg:1979pi}
\citentry{Salam:1980jd}
\citentry{Glashow:1979pj}

Experiments (and the supporting theoretical calculations) over the 
past decade have elevated the electroweak theory to a law of nature. 
The current state of the theory is reviewed in 
\citentry{Quigg:2009vq}

For general surveys of the standard model of particle physics, and a
glimpse beyond, see
\citentry{Gaillard:1998ui}
\citentry{Rosner:2002xi}

The common mathematical structure of QCD and the electroweak theory,
combined with asymptotic freedom, encourages the hope that a unified
theory of the strong, weak, and electromagnetic interactions may be
within reach.
The unification strategy, with some consequences, is presented in
\citentry{Georgi:1981yp}

\section{Frontier Problems in QCD}
\label{sec:outlook}

Four decades after the synthesis of quarks, partons, and color into the 
QCD Lagrangian (Ref.~\onlinecite{Fritzsch:1973pi})---and the essentially 
immediate discovery of asymptotic 
freedom (Ref.~\onlinecite{Gross:1973id,Politzer:1973fx})---QCD 
has been tested and validated up to energies of 1~TeV.
Tests are poised to continue at even higher energies, as operations at 
the Large Hadron Collider (LHC) commence.
It is fair to say, however, that most physicists do not expect big 
surprises at the LHC in the structure of QCD.
Instead, QCD will be treated as basic knowledge, much like 
electrodynamics, enabling discoveries beyond the realm of the standard  
model of elementary particles (Ref.~\onlinecite{Ellis:1991qj}).

In this arena, future research will focus on techniques for evaluating 
parton amplitudes with increasingly many real and virtual particles,
for both signals and backgrounds.
The higher energies of the scattering processes will continue to entail 
many scales (several TeV compared to the top-quark mass, for example) 
and, hence, will need tools, such as the soft-collinear effective theory
discussed in Sec.~\ref{sec:scet}.
Future experiments with $B$ decays will also continue to rely on QCD, 
at moderately high energies,  to pin down the weak and any new 
interactions of quarks (or other particles carrying color).

The strong interactions comprise a richer field than the set of 
phenomena that we have learned to describe in terms of perturbative QCD 
or the (near-)static nonperturbative domain of lattice QCD.
The technology by which we apply QCD is incomplete, and still evolving.
Many aspects of hadron phenomenology and spectroscopy are not yet 
calculable beginning from the QCD Lagrangian.
Much analysis of experimental information relies on highly stylized, 
truncated pictures of the implications of the theory.
While expanding the horizons, it is important to distinguish tests of 
QCD from tests of auxiliary assumptions.

The rest of the strong interactions, moreover, isn't confined to common  
processes with large cross sections such as the ``soft'' particle 
production, elastic scattering, or diffraction.
It may well be that interesting, \textit{unusual} occurrences happen 
outside the framework of perturbative QCD---happen in some collective, 
or intrinsically nonperturbative, way.
At the highest energies, well into the regime where the $pp$ total 
cross section grows as $\ln^2s$, long-range correlations might show 
themselves in new ways.
Quantum chromodynamics suggests new, modestly collective, effects such 
as multiple-parton interactions.
The high density of partons carrying $p_z=5$--$10\gev$ may give rise to 
hot spots in the spacetime evolution of the collision aftermath, and 
thus to thermalization or other phenomena not easy to anticipate from 
the QCD Lagrangian.

At lower energies, the basic features of the hadron spectrum have been 
reproduced in a convincing way.
Some of the simplest hadronic transition amplitudes, needed to understand 
flavor physics, are in similarly good shape.
The aspiration here is to compute many simple amplitudes with total errors 
that are 1\% or smaller.
Such precision will require nonperturbative matching and the charmed 
sea.
Indeed, a next-generation assault on $B$ decays via 
$e^+e^-\to\Upsilon(\mathrm{4S})$ will hinge on such lattice QCD 
calculations (Ref.~\onlinecite{Antonelli:2009ws}).
Calculations of similar difficulty are related to moments of the parton 
distributions.
Reliable lattice-QCD calculations would pin down predictions of signals 
and backgrounds at the LHC.
The most crucial in this regard, and most challenging computationally, 
are moments of the gluon density inside the proton.
See Ref.~\onlinecite{Hagler:2009ni} and
\citentry{Renner:2010ks}

Precision perturbative QCD and precision lattice QCD are important and 
challenging, yet programmatic.
Other future avenues for research in QCD will explore its richness in 
ways that are harder to anticipate.
QCD is frequently, and justifiably, hailed as a triumph of reductionist 
science, distilling the plethora of hadrons and their complicated 
properties into a simple Lagrangian field theory [\Eqn{eq:qcdlag}].
Now that QCD is accepted as a law of nature, however, it may be time to 
characterize QCD research by the phenomena that emerge from 
this tantalizing simple form.
What are hadron masses and chiral symmetry breaking, if not emergent 
phenomena?

Many avenues offer themselves for quantitative and qualitative study.
Although the spectrum of the lowest-lying conventional hadrons is 
well-computed, it remains a challenge to compute the masses of excited 
hadrons, and even the lowest-lying glueball, hybrid, and exotic states.
While these masses tie into experimental programs, it would simply be 
intriguing to see towers of bound states emerge from the QCD Lagrangian.
Another structure that emerges from QCD is a rich phase structure
(see Sec.~\ref{sec:thermo}). 
A fuller understanding will require experiments with heavy-ion collisions, 
including the higher-density probes of the Compressed Baryonic Matter 
experiment.
\citentry{CBM}
\noindent Complementary theoretical work will require both model studies and 
lattice QCD calculations, although a breakthrough in finite-density 
lattice QCD could relegate some model studies to secondary importance.
The transition to (effectively) deconfined quarks at nonzero 
temperature and density may help explain why color cannot be 
isolated in the (zero-temperature) ground state of QCD.
Finally, from the emergent phenomenon of hadrons emerges the whole 
field of nuclear physics.
QCD is just beginning to answer questions about nuclear physics, and 
some nuclear physicists see the future of their field as QCD
(see Sec.~\ref{sec:nuke}).

To elucidate these features of QCD, it will help to study lightweight 
versions of Yang-Mills theories with quarks.
For example, with one flavor there is no chiral symmetry to break---the 
anomaly represents an explicit breaking of the U(1) chiral symmetry
(see Sec.~\ref{sec:anomaly})---presenting a laboratory to study 
confinement without spontaneous symmetry breaking:
\citentry{Farchioni:2007dw}
\noindent What properties does this confining theory share with QCD?
What does it lose along with the loss of chiral symmetry, spontaneously 
broken?
An irony of nature's version of QCD is that the up- and down-quark 
masses are so much smaller than $\Lambda_{\mathrm{QCD}}$, so isospin is 
an excellent approximate symmetry.
(More properly, isospin follows from $m_d-m_u\ll\Lambda_{\mathrm{QCD}}$.)
Alternatively, one could imagine a theory with two quarks whose masses, 
and mass difference, are comparable to or larger 
than~$\Lambda_{\mathrm{QCD}}$.
Which dynamical features remain, and which are lost?

Whatever results academic investigations bring, QCD will retain a 
strong and deep connection to particle physics, astrophysics and 
cosmology, and nuclear physics.
Indeed, often QCD binds these fields to each other.
As discussed above, QCD will always play a central role, within the 
standard model and beyond, for collider physics.
A dream of particle theorists is to unify the strong, weak, and 
electromagnetic interactions.
Further precision for $\alphas$ and quark masses will inform and 
constrain this dream.
Now that it is fairly well established that the up-quark mass cannot 
vanish, the strong CP problem demands other solutions.
The most elegant proposal augments QCD with additional 
symmetry (Ref.~\onlinecite{Peccei:1977hh}).
The observable consequence is a pseudoscalar particle called the axion, 
which may comprise part of the ``dark'' matter of the 
universe (Ref.~\onlinecite{Kim:2008hd}).

A future challenge is to connect nuclei to QCD.
Some aspects befuddle models, because some relevant properties are too 
hard to measure.
For example, the three-nucleon interaction is an important missing 
piece to the puzzle of nuclear structure.
Some questions are almost philosophical:
How do $\alphas$ and the quark masses lead to various 
happenstances of nuclear physics, some of which seem implausible, yet 
are necessary for carbon-based life to exist?
Looking beyond Earth, details of the quark-gluon plasma influence the 
evolution of the early universe.
Above the transition temperature, hadrons do not ``dissolve'' quite as 
fast as sometimes thought.
Another interesting QCD calculation, which has not yet been carried out, 
is to determine the $\Sigma^-$-nucleon interaction.
This is not a prosaic matter of hadronic physics, but a nuclear 
property that influences whether a supernova evolves to a neutron star 
or a black hole (Ref.~\onlinecite{Beane:2008dv}).

In summary, QCD is not our ``most perfect theory'' 
(Ref.~\onlinecite{Wilczek:1999id}) merely because asymptotic freedom 
ensures its scope on towards the highest energies, temperatures, and 
densities.
It is also a rich and varied physical theory, exhibiting qualitatively 
different behavior in different regimes, all stemming ultimately on the 
dynamics of quarks and gluons.
For all the explanatory power of QCD, it still provides problems for 
physicists to work~on.

\acknowledgments

We thank 
Thomas Becher,
Johan Bijnens,
Lance Dixon,
Claudia Glasman,
Maxime Gouzevitch,
Kenichi Hatekeyama,
Frithjof Karsch,
Stefan Kluth,
Christina Mesropian,
and
Heath O'Connell
for assistance in the preparation of this Resource Letter.

Fermilab is operated by Fermi Research Alliance, LLC  under Contract
No.~DE-AC02-07CH11359 with the United States Department of Energy.  

\appendix

\section{Links to Basic Resources}
\label{app:links}

\subsection{Journals}
New research papers on QCD are published in journals of elementary 
particle physics and of nuclear physics.
The principal particle physics journals are
\begin{itemize}
	\item \emph{European Physics Journal C}, available on-line at 
	\url{http://epjc.edpsciences.org/};
	\item \emph{Journal of High Energy Physics}, available on-line at 
	\url{http://jhep.sissa.it/jhep/} or 
	\url{http://www.iop.org/EJ/jhep/};
	\item \emph{Journal of Physics G}, available on-line at 
	\url{http://www.iop.org/EJ/journal/JPhysG/};
	\item \emph{Nuclear Physics B}, available on-line at 
	\url{http://www.elsevier.com/wps/product/cws_home/505716/};
	\item \emph{Physical Review D}, available on-line at 
	\url{http://prd.aps.org/};
	\item \emph{Physical Review Letters}, available on-line at 
	\url{http://prl.aps.org/};
	\item \emph{Physics Letters B}, available on-line at 
	\url{http://www.elsevier.com/wps/product/cws_home/505706/}.
\end{itemize}
The principal nuclear physics journals are
\begin{itemize}
	\item \emph{European Physics Journal A}, available on-line at 
	\url{http://epja.edpsciences.org/};
	\item \emph{Journal of Physics G}, available on-line at 
	\url{http://www.iop.org/EJ/journal/JPhysG/};
	\item \emph{Nuclear Physics A}, available on-line at 
	\url{http://www.elsevier.com/wps/product/cws_home/505715/};
	\item \emph{Physical Review C}, available on-line at 
	\url{http://prc.aps.org/};
	\item \emph{Physical Review Letters}, available on-line at 
	\url{http://prl.aps.org/};
	\item \emph{Physics Letters B}, available on-line at 
	\url{http://www.elsevier.com/wps/product/cws_home/505706/}.
\end{itemize}
Journals with review articles:
\begin{itemize}
	\item \emph{Annual Reviews of Nuclear and Particle Science}, 
	available on-line at 
	\url{http://arjournals.annualreviews.org/loi/nucl}
	\item \emph{Physics Reports}, available on-line at 
	\url{http://www.elsevier.com/wps/product/cws_home/505703/}
	\item \emph{Reports on Progress in Physics}, available on-line at 
	\url{http://www.iop.org/EJ/journal/RoPP/}
	\item \emph{Reviews of Modern Physics}, available on-line at 
	\url{http://rmp.aps.org/}
\end{itemize}

These websites provide electronic versions (\eg, pdf files) of
most---in some cases all---papers published in the corresponding journal.
Often a personal or institutional subscription, or the payment of a fee, 
is necessary.

\subsection{Electronic archives}
Most research papers and conference proceedings appear first in the 
physics e-print archives:
\begin{itemize}
	\item \url{http://arxiv.org/archive/hep-ex/} contains e-prints on 
	experimental high-energy (elementary particle) physics, many of 
	which concern QCD;
	\item \url{http://arxiv.org/archive/hep-lat/} contains e-prints on 
	lattice gauge theory, most of which address nonperturbative QCD;
	\item \url{http://arxiv.org/archive/hep-ph/} contains e-prints on 
	theoretical high-energy physics with focus on observable phenomena, 
	many of which concern QCD;
	\item \url{http://arxiv.org/archive/hep-th/} contains e-prints on 
	theoretical aspects of string theory and quantum field theory,
	some of which concern QCD;
	\item \url{http://arxiv.org/archive/nucl-ex/} contains e-prints on 
	experimental nuclear physics, many of which concern QCD explicitly;
	\item \url{http://arxiv.org/archive/nucl-th/} contains e-prints on 
	theoretical nuclear physics, many of which concern QCD explicitly.
\end{itemize}
The arXiv provides free downloads.
The arXiv version of this Resource Letter provides hyperlinks to 
\texttt{arXiv.org} where possible, and otherwise provides a hyperlink
to the digital object identifier (doi) of other electronically 
published sources.
One should bear in mind, however, that the versions in journals are 
usually definitive; \texttt{arXiv.org} provides doi links.

\subsection{Pedagogical web sites}

For a very approachable introduction to the ideas of contemporary
particle physics, see the
\citentry{CPEPchart}
\noindent and the accompanying
\citentry{CPEPbook}

The ideas of nuclear science are presented in a wall chart and 
teacher's guide, available at
\citentry{CPEPnuc}

The Particle Data Group (Ref.~\onlinecite{Amsler:2008zzb}) maintains a 
web site as comprehensive as its review, with updates online midway 
between the biennial editions.
Students and the general public should enjoy their Particle Adventure,
\url{http://www.particleadventure.org/}.

Visualizations of some of the main elements of nonperturbative 
QCD, with helpful explanations, may be found at the URL 
in Ref.~\onlinecite{visqcd}.

The laboratories and major experiments in particle and nuclear physics 
maintain web sites that feature educational materials.

\end{document}